\documentclass[10pt,a4paper,final]{iopart}      

\usepackage{iopams}
\usepackage{amstext}
\usepackage{graphicx}
\usepackage{pstricks}
\usepackage[breaklinks=true,colorlinks=true,linkcolor=blue,urlcolor=blue,citecolor=blue]{hyperref}
\usepackage{cite}
\hypersetup{%
pdfstartview={XYZ null null 1.25}
}

\newcommand{\ep}{$e$-$ph$~}
\newcommand{\pdag}{{\phantom\dagger}}
\newcommand{\bk}{{\bf k}}
\newcommand{\bq}{{\bf q}}
\newcommand{\bp}{{\bf p}}

\def\ioptwocol{\setlength\hoffset{-0.5in}\setlength\voffset{-0.5in}\setlength\textwidth{6.75in}
\setlength\columnsep{0.2in}\setlength\textheight{9.25in}\mathindent=0in\twocolumn}

\graphicspath{{../figures/latex_fig/}{../figures/}}

\begin{document}

\title[Aspects of \ep interactions with strong forward scattering in FeSe/STO]
{Aspects of electron-phonon interactions with strong forward scattering in FeSe Thin Films on SrTiO$_3$ substrates}

\author{Y Wang$^1$, K Nakatsukasa$^1$, L Rademaker$^2$, T Berlijn$^{3,4}$ and S Johnston$^1$}
\address{$^1$ Department of Physics and Astronomy, University of Tennessee, Knoxville, Tennessee 37996, USA}
\address{$^2$ Kavli Institute for Theoretical Physics, University of California Santa Barbara, California 93106, USA}
\address{$^3$ Center for Nanophase Materials Sciences, Oak Ridge National Laboratory, Oak Ridge, Tennessee 37831, USA}
\address{$^4$ Computer Science and Mathematics Division, Oak Ridge National Laboratory, Oak Ridge, Tennessee 37831, USA}
\eads{\mailto{sjohn145@utk.edu}}

\begin{abstract}
Mono- and multilayer FeSe thin films grown on SrTiO$_\mathrm{3}$ and BiTiO$_\mathrm{3}$ substrates exhibit
a greatly enhanced superconductivity over that found in bulk FeSe. A number of proposals have been advanced
for the mechanism of this enhancement. One possibility is the introduction of a cross-interface
electron-phonon ($e$-$ph$) interaction between the FeSe electrons and oxygen phonons in the substrates that
is peaked in the forward scattering (small $\bq)$ direction due to the two-dimensional nature of the
interface system. Motivated by this, we explore the consequences of such an interaction on the
superconducting state and electronic structure of a two-dimensional system using Migdal-Eliashberg theory.
This interaction produces not only deviations from the expectations of conventional phonon-mediated pairing
but also replica structures in the spectral function and density of states, as probed by angle-resolved
photoemission spectroscopy, scanning tunneling microscopy/spectroscopy, and quasi-particle interference
imaging. We also discuss the applicability of Migdal-Eliashberg theory for a situation where the \ep
interaction is peaked at small momentum transfer and in the FeSe/STO system.
\end{abstract}

\pacs{71.38.-k,74.10.+v,63.22.-m, 74.70.Xa}\submitto{\SUST}
\vspace{2pc} \noindent{\it Keywords\/}: Superconductivity, Forward Scattering, Eliashberg Theory, FeSe films
\maketitle
\ioptwocol

\section{Introduction}

Bulk iron selenide (FeSe) is an unconventional superconductor with a relatively
modest transition temperature $T_\mathrm{c} \sim 9$ K at ambient pressures
\cite{BulkFeSe}; however, its $T_\mathrm{c}$ can be increased by upwards of a
factor of ten when a monolayer of FeSe is deposited on SrTiO$_3$ (STO)
\cite{WangCPL2012} or BaTiO$_3$ (BTO) substrates \cite{PengNatureComm2014}.
This discovery has attracted considerable scientific interest
\cite{WangCPL2012,PengNatureComm2014,LiuNatureComm2012,ZhangPRB2014,
ZhengSciRep2013,PengPRL2014,LiuNatureComm2014,LiuPRB2012,LeeNature2014,
LeePreprint,RademakerPreprint,GeNature,HeNatureMaterials2013,MiyataNatureMat2015,
TanNatureMaterials2013,LiAPL2014,CohNJP2015,HuangPRL2015,FanPreprint,
SongPreprint,SeoPreprint,Shiogai, ZhangPreprint,ZhouPreprint,
GapAnisotropy,LeeQM} as it provides
not only a new route to high-temperature (high-$T_\mathrm{c}$) superconductivity
but also a new lens through which unconventional superconductivity can be studied.

Several proposals have been advanced for the microscopic origin of this
enhancement and the most widely discussed scenarios are broadly divided into
two categories. The first category involves charge transfer between the
substrate and the film, which dopes the monolayer with excess electrons. This
may result in modifications of the electronic structure, which
in turn enhances an unconventional pairing mechanism \cite{LiuNatureComm2012}
or suppresses competing phases \cite{TanNatureMaterials2013}.
In addition to this, the charge transfer shifts
the holelike bands centered at $\Gamma = (0,0)$ below the Fermi
level \cite{LiuNatureComm2012,HeNatureMaterials2013,TanNatureMaterials2013,
LeeNature2014}, creating an electronic structure similar to the
intercalated FeSe systems \cite{DagottoRMP,Intercalated1,Intercalated3}.
This electronic structure challenges the
Fermi-surface-nesting driven, purely electronic pairing mechanism common to the
iron-based superconductors \cite{HirschfeldReview,MazinPRL}.  The second
category of proposals for FeSe/STO encompasses interface-related effects, where
a more direct role is played by the substrate. The main proposals here include the
suppression of a competing phase via strain
 or modifications of the
electron-phonon ($e$-$ph$) interaction in the FeSe layer \cite{CohNJP2015}, or
the introduction of one across the interface \cite{LeeNature2014,LeePreprint}.

In this paper we focus on the last scenario, which is motivated by several key experimental observations. The
first is that the intercalated FeSe systems \cite{Intercalated1,Intercalated2,Intercalated3} and FeSe/STO
thin films post-treated with K~\cite{MiyataNatureMat2015} or Na~\cite{SeoPreprint} adatoms all have maximum
$T_\mathrm{c}$ values $\sim 40$~K. This is despite the fact that all of these systems have a similar
electronic structure to FeSe monolayers on STO.  This indicates that any common unconventional pairing
mechanism in these systems is insufficient to account for the observed high-$T_\mathrm{c}\sim
55\text{--}75$~K in monolayer FeSe/STO and FeSe/BTO and an additional contribution to pairing is likely
needed \cite{LeePreprint}. The second (and more important) observation is that of replica bands in the
electronic structure of superconducting FeSe/STO \cite{LeeNature2014,PengNatureComm2014}. These are
interpreted as shake-off states arise from a coupling between the FeSe 3$d$ electrons and an optical O bond
stretching phonon branch in the oxide substrate. The crucial observation here is the size and shape of these
replica bands---being complete copies of the main electron band--- shows that the \ep coupling constant must
be strongly peaked in the forward scattering direction (small $\bq$).  Because of this momentum structure,
the cross-interface coupling is strongly intraband in nature and can contribute to pairing in most channels,
even those commonly associated with repulsive pairing mechanisms relevant for the iron-based superconductors
\cite{FS1,FS2,FS3,FS4,BulutPRB1996,ChenPreprint,LeeQM}. These observations make the cross-interface coupling
a likely candidate for the additional pairing needed to produce the observed $T_\mathrm{c}$ in FeSe/STO
\cite{LeeNature2014,RademakerPreprint, LeePreprint,LeeQM}.

As this is a special issue on {\em superconductivity in the two-dimensional (2D) limit} it is worth stressing
that the forward focus of the cross-interface \ep interaction can be understood as a 
consequence of dimensionality of the FeSe
thin films and the anisotropic dielectric properties of the interface.  A more complete discussion of this
aspect can be found in Ref. \cite{LeePreprint}.  Here, we summarize the main points. The FeSe films sit a
distance $h$ above the TiO terminated STO substrate \cite{WangCPL2012}. The motion of the oxygen atoms in the
direction perpendicular to the FeSe film induce a local dipole moment with an effective charge
$q_\mathrm{eff}$ at the surface of the substrate. As a result, the optical O modes 
create an electric dipole potential on the FeSe electrons. In the continuum limit, this
perturbation will dynamically scatter the FeSe electrons from states with in-plane momentum $\bk$ to $\bk +
\bq$ with a matrix element
\begin{equation}\label{Eq:Gq}
g(\bq) = \frac{2\pi q_\mathrm{eff}}{\epsilon_\perp}\exp\left(-\frac{|\bq|}{q_0}\right).
\end{equation}
Here, $\bq$ is the in-plane momentum transfer, $q_0^{-1} = h\sqrt{\frac{\epsilon_\parallel}{\epsilon_\perp}}$
sets the range of the interaction in momentum space, and $\epsilon_\parallel$ ($\epsilon_\perp$) is the
dielectric constant parallel (perpendicular) to the interface. In FeSe/STO (and also FeSe/BTO) one expects
$\epsilon_\parallel \gg \epsilon_\perp$ \cite{LeePreprint}, since the FeSe film can provide additional
metallic screening parallel to the interface. In this limit, $q_0$ is small, resulting in an  
interaction that is peaked at $\bq = 0$.

Previously, some of the present authors showed that a strong forward-focused interaction not only reproduces
the replica band structure but also acts as a remarkably effective pairing mediator with a superconducting
$T_\mathrm{c} \propto \lambda_\mathrm{m}$ in the weak coupling limit \cite{RademakerPreprint}. As such, the
cross-interface \ep coupling implied by the replica bands can provide a substantial contribution to
$T_\mathrm{c}$ and this can work in conjunction with an unconventional pairing mechanism. This outlook is
supported by recent projector quantum Monte Carlo results \cite{LeeQM} and multiband BCS results involving
the incipient hole bands at the $\Gamma$-point \cite{ChenPreprint}. It is therefore important to understand
and confirm the exact role played by the cross-interface coupling. Motivated by this, we build upon our
previous work \cite{RademakerPreprint} and present several aspects of a 2D system of electrons coupled to an
optical phonon branch by forward focused interaction. Our aim is to draw a more complete picture of the
consequences of such interaction and to provide further spectroscopic signatures that can be used to confirm
its presence. We note that while our focus is on the FeSe/STO system, similar forward
scattering may occur in other low-dimensional correlated systems \cite{FS2,FS3,FS4,BulutPRB1996}. We hope
that our results will provide additional means to confirm or rule out this possibility.

The paper is organized as follows. Section~\ref{Sec:Formalism} provides the details of the model and
Migdal-Eliashberg formalism used throughout the paper. Section~\ref{Sec:Migdal} discusses the validity of
this approach in FeSe/STO, which has been questioned recently \cite{GorkovPreprint}. Our major results are
presented in section~\ref{Sec:Results}, beginning in section~\ref{Sec:Akw}, where we review our prior
results for the formation of the replica bands and present new details. In section~\ref{Sec:Gap} we then examine
the properties of the superconducting state and the anisotropy of the superconducting order parameter.
Finally, in sections~\ref{Sec:DOS} and \ref{Sec:QPI} we present results for electronic density of states
and quasiparticle interference patterns, respectively.  Finally, section~\ref{Sec:Conclusions}
provides additional discussion and concluding remarks.

\section{Formalism}\label{Sec:Formalism}
\subsection{Model Hamiltonian}
We consider a simplified single-band model for FeSe electrons coupled to an optical phonon branch via a
strongly momentum-dependent interaction. The system is described by the Hamiltonian
\begin{eqnarray*}
\fl H =& \sum_{\bk,\sigma} \xi^\pdag_\bk c^\dagger_{\bk,\sigma}c^\pdag_{\bk,\sigma}
  + \sum_\bq \Omega^\pdag_\bq \left(b^\dagger_\bq b^\pdag_\bq+\frac{1}{2}\right) \\
  &+ \frac{1}{\sqrt{N}}\sum_{\bk,\bq,\sigma} g(\bk,\bq) c^\dagger_{\bk+\bq,\sigma} c^\pdag_{\bk,\sigma}
  \left(b^\dagger_{-\bq} + b^\pdag_\bq \right),
\end{eqnarray*}
where $c^\dagger_{\bk,\sigma}$ ($c^\pdag_{\bk,\sigma}$) creates (annihilates) an electron with wavevector
$\bk$ and spin $\sigma$, $b^\dagger_\bq$ ($b^\pdag_\bq$) creates (annihilates) a phonon with wavevector
$\bq$; $\xi_\bk$ is the electronic band dispersion measured relative to the chemical potential $\mu$;
$\Omega_\bq$ is the phonon dispersion ($\hbar = 1$); and $g(\bk,\bq)$ is the momentum dependent \ep
interaction.

The electronic band dispersion has a simple form $\xi_\bk = -2t[\cos(k_xa)+\cos(k_ya)] - \mu$, where $a$ is
the in-plane lattice constant. As we are primarily motivated by the case of FeSe/STO, we set $t = 75$ meV and
$\mu = -235$ meV unless otherwise stated. This parameter set produces a single electron-like Fermi surface
with $k_\mathrm{F} = 0.97/a$ and a Fermi velocity $v_\mathrm{F} = 0.12$~eV$\cdot a/\hbar$ along the $k_y = 0$
line, which is similar to the electron pocket at the $M$ point in FeSe monolayers on STO
\cite{LiuNatureComm2012,HeNatureMaterials2013,TanNatureMaterials2013,LeeNature2014}. We approximate the
relevant phonon branch with a dispersionless $\Omega_\bq = \Omega = 100$ meV optical oxygen phonon mode. This
is supported by several {\it ab initio} calculations that indicate that the highest lying phonon branch in
bulk STO \cite{PhononsSTO} and FeSe/STO interfaces \cite{PhononsFeSeSTO,WangDFT} is relatively flat in the
vicinity of $\bq = 0$. The \ep matrix element is taken to depend on the momentum transfer only, as derived
microscopically in previous works \cite{LeeNature2014,LeePreprint}. Specifically, we parameterize the
coupling as $g(\bq) = g_0\exp(-|\bq|/q_0)$, where $g_0$ is adjusted to fix the total dimensionless coupling
strength of the interaction (see section \ref{Sec:lambda}) and $q_0$ sets the range of the interaction in
momentum space. Previously, we showed that $q_0$ must be small in order to reproduce the spectral properties
of the replica band \cite{LeeNature2014,RademakerPreprint} (see also Fig.~\ref{Fig:Akwlmd}). This implies
that the \ep coupling is strongly peaked in the forward scattering direction (small $|\bq|$).

\subsection{Migdal-Eliashberg Theory}

\begin{figure*}
  \centering
  \includegraphics[width=0.7\textwidth]{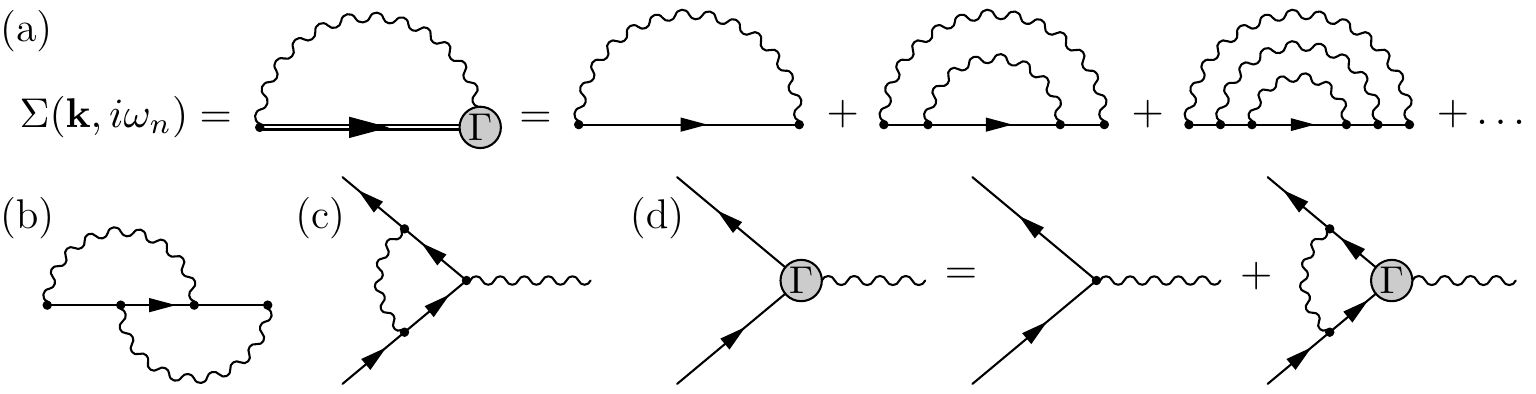}

\caption{\label{Fig:Diagrams}The Feynman diagrams relevant for our discussion of the electron-phonon
self-energy. (a) The set of rainbow diagrams summed to obtain the self-energy $\Sigma(\bk,i\omega_n)$ in the
Migdal-Eliashberg formalism. (b) The lowest order crossing diagram neglected in the same formalism. (c) The
diagram for the lowest order vertex correction $\Gamma^{(1)}(i\omega_n,i\omega_{n^\prime})$. 
(d) The summation of the vertex diagrams within the ladder approximation. }
\end{figure*}

We study the model Hamiltonian using Migdal-Eliashberg (ME) theory. (The validity of this approach for the
FeSe/STO system is discussed further in section \ref{Sec:Migdal}.) With the Nambu notation the electronic
self-energy in the superconducting state and on the Matsubara frequency axis can be partitioned as
\begin{eqnarray}
\fl \hat{\Sigma}(\bk,i\omega_n) =& i\omega_n[1-Z({\bf k},i\omega_n)]\hat{\tau}_0
         + \chi(\bk,i\omega_n)\hat{\tau}_3  \nonumber\\
        &+ \phi(\bk,i\omega_n)\hat{\tau}_1,
\end{eqnarray}
where $\hat{\tau}_\alpha$ are the Pauli matrices, $Z(\bk,i\omega_n)$ is the quasiparticle weight,
$\chi(\bk,i\omega_n)$ renormalizes the band dispersion, and $\phi(\bk,i\omega_n)$ is the anomalous
self-energy (which is zero in the normal state).

In ME theory, the self-energy is obtained by summing the
single-loop diagrams shown in Fig.~\ref{Fig:Diagrams}(a). This leads to a self-consistency equation
\begin{eqnarray*}
\fl \hat{\Sigma}(\bk,i\omega_n) =& -\frac{1}{N\beta}\sum_{\bq,m}
        |g(\bk,\bq)|^2D^{(0)}(\bq,i\omega_n-i\omega_m) \nonumber\\
    &\times \hat{\tau}_3 \hat{G}(\bk+\bq,i\omega_m)\hat{\tau}_3.
\end{eqnarray*}
Here, $\hat{G}^{-1}({\bf k},i\omega_n) = i\omega_n\hat{\tau}_0 - \xi_{\bf k}\hat{\tau_3}
-\hat{\Sigma}(\bk,i\omega_n)$ and $D^{(0)}(\bq,i\omega_\nu) = -\frac{2\Omega_\bq}{\Omega^2_\bq +
\omega_\nu^2}$ are the dressed electron and bare phonon Green's functions, respectively; $\omega_n =
\pi(2n+1)/\beta$ and $\omega_\nu = 2\nu\pi/\beta$ are fermionic and bosonic Matsubara frequencies,
respectively; and $\beta = 1/T$ ($k_\text{B} = 1$) is the inverse temperature.

The self-energy on the real frequency axis is obtained from the analytic continuation of the imaginary axis
solutions with $i\omega_n \rightarrow \omega + i\eta$. We carry out this procedure using the efficient
two-step iterative procedure given in Ref. \cite{MarsiglioPRB1989}, where the self-energy is obtained from
self-consistently solving
\begin{eqnarray}\label{Eq:SE}
\fl \hat{\Sigma}({\bf k},\omega + i\eta) = \nonumber\\
\quad   -\frac{1}{N\beta}\sum_{m,\bq} \int dz |g(\bk,\bq)|^2 B(z)
        \frac{\hat{\tau}_3\hat{G}(\bk+\bq,i\omega_m)\hat{\tau}_3}{\omega-i\omega_m - z} \nonumber\\
\quad    + \frac{1}{N}\sum_\bq \int dz |g(\bk,\bq)|^2 B(z)
        \hat{\tau}_3\hat{G}(\bk+\bq,\omega-z+i\eta)\hat{\tau}_3 \nonumber\\
\quad    \times\frac{1}{2}\left[\tanh\left(\frac{(\omega-z)\beta}{2}\right)
        + \coth\left(\frac{\beta z}{2}\right)\right].
\end{eqnarray}
In this case, the \ep coupling constant depends on the momenutum transfer $\bq$ only.
The momentum summations
can therefore be evaluated efficiently using fast Fourier transforms.

\subsection{Dimensionless Coupling}\label{Sec:lambda}
\begin{figure}[t]
  \centering
  \includegraphics[width=0.8\columnwidth]{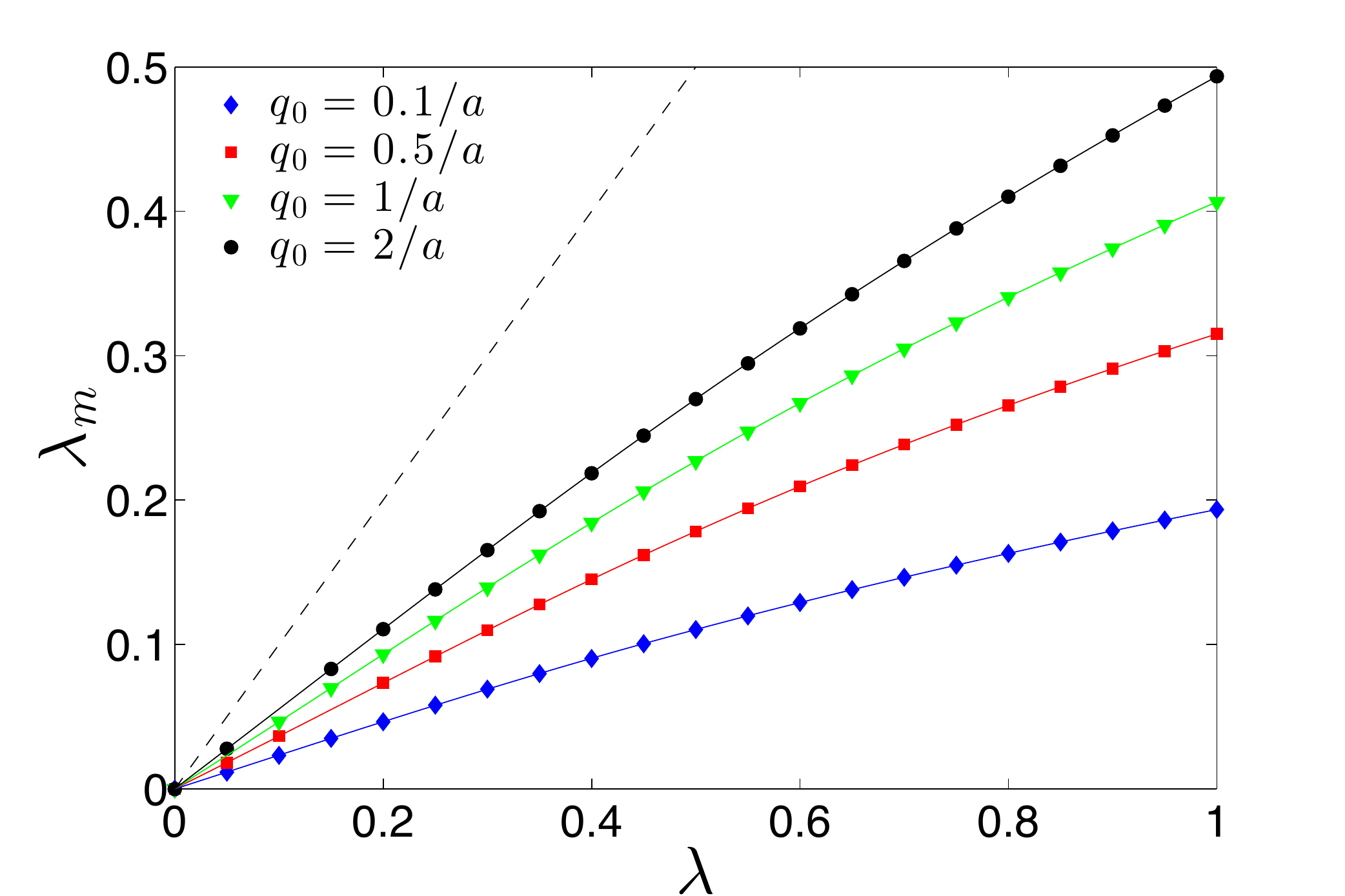}

\caption{\label{Fig:lambda}The discrepancy of the average value of mass enhancement $\lambda_\mathrm{m}$ and the
conventional definition for the dimensionless coupling $\lambda$ obtained from our numerical calculations at
$T = 100$ K. These results are obtained on a $128\times128$ site lattice and with Gaussian broadening of 15
meV. The dashed line indicates the line for $\lambda_\mathrm{m} = \lambda$ as would be expected for a
momentum-independent coupling.}
\end{figure}

It is common practice to define a dimensionless measure of the interaction strength $\lambda_\mathrm{m}$ when studying
electron-boson coupled systems. Physically, this value characterizes the average value of the quasiparticle
mass renormalization on the Fermi surface due to the interaction. It is defined as $\lambda_\mathrm{m} = \sum_\bk
\lambda_\bk \delta(\xi_\bk)/\sum_\bk \delta(\xi_\bk)$, where
\begin{equation}
\lambda_\bk = -\mathrm{Re}\frac{\partial \Sigma(\bk,\omega)}{\partial\omega}\bigg|_{\omega = 0}
= {\mathrm{Re}}[Z(\bk,\omega=0)-1].
\end{equation}
An alternative measure of this coupling, denoted $\lambda$, is commonly employed in the
literature. It is given by a
double Fermi-surface-average of the \ep coupling constant
\begin{equation}\label{Eq:LambdaWrong}
\lambda \equiv \frac{2}{N_\mathrm{F}\Omega N^2} \sum_{\bk,\bk^\prime}
|g(\bk,\bk^\prime)|^2 \delta(\xi_\bk)\delta(\xi_{\bk^\prime}).
\end{equation}

In the context of this work it is important to stress that these two definitions are \emph{not} equivalent;
only when the momentum dependence of $g(\bk,\bq)$ is weak and $\Omega \ll E_\mathrm{F}$ does $\lambda_\mathrm{m}
\approx \lambda$ \cite{Mahan}. This is illustrated for our forward scattering interaction in
Fig.~\ref{Fig:lambda}, which plots the physical $\lambda_\mathrm{m}$ vs. the common definition $\lambda$ for various
values of $q_0$. One sees clearly that $\lambda_\mathrm{m}$ differs significantly from $\lambda$ and the size of
the discrepancy grows as the interaction becomes increasingly peaked at $\bq = 0$. This indicates that a
strongly forward-focused \ep interaction is less effective at dressing electrons, resulting in dressed
carriers with lighter effective masses. This is one factor contributing to the large $T_\mathrm{c}$ values obtained
from this interaction.
For the remainder of this work we use $\lambda_\mathrm{m}$ to characterize the strength of the \ep coupling.

\subsection{Migdal's Theorem with Strong Forward Scattering}\label{Sec:Migdal}
In ME theory the self-energy is computed by summing the diagrams shown in Fig.~\ref{Fig:Diagrams}(a), while the
crossing diagrams like the one shown in Fig.~\ref{Fig:Diagrams}(b) [involving vertex corrections like
the one shown in Fig.~\ref{Fig:Diagrams}(c)] are neglected. Justification for this approximation is typically
provided by Migdal's theorem, which states that the vertex corrections in a metallic system with a wide
bandwidth
are proportional to the product of the dimensionless coupling $\lambda_\mathrm{m}$ times the adiabatic ratio
$\Omega/E_\text{F}$ \cite{Migdal}. Here, $\Omega$ parameterizes the typical energy of the relevant phonons
and $E_\text{F}$ is the Fermi energy
of the electron.
In a metallic system the lowest order vertex correction is proportional to
$\lambda_\mathrm{m}\frac{\Omega}{E_\text{F}} \propto \sqrt{\frac{m}{M}}$, where $m$ and
$M$ are the electron and ion masses, respectively. Thus the self-energy corrections due to the crossing
diagrams are much smaller than the rainbow diagrams and can be neglected at all orders.

The validity of this approach has been questioned \cite{GorkovPreprint} in FeSe/STO, where $\Omega \sim 100$
meV and $E_\mathrm{F} \sim 65$ meV. It turns out, however, that the strong forward-scattering peak in the
interaction removes the adiabatic ratio from the prefactor of the vertex correction. As a result, Migdal's
theorem has the potential to break down in {\it any} system dominated by small $\bq$ scattering, regardless
of the value of $E_\mathrm{F}$. To demonstrate this, we consider the lowest order vertex correction
$\Gamma^{(1)}_{\bk,\bk^\prime}(\omega_n,\omega_{n^\prime})$  [Fig.~\ref{Fig:Diagrams}(c)] in the limit of
perfect forward scattering, where $|g(\bq)|^2 = g_0^2 N\delta_{\bq,0}$\cite{RademakerPreprint}. In this
limit, no momentum is transferred to the electron and we can set the external momenta to be 
equal $\bk=\bk^\prime$. The vertex correction is then given by

\begin{eqnarray*}
\fl \Gamma_\bk^{(1)}(\omega_n,\omega_{n^\prime}) =&\\
-\frac{g_0^2}{\beta}\sum_m D^{(0)}(\omega_m)G^{(0)}(\bk,\omega_{n^\prime}-\omega_m)G^{(0)}(\bk,\omega_n-\omega_m) \\
=\frac{g_0^2}{\beta}\sum_m \frac{2\Omega}{\Omega^2 + \omega_m^2} 
\frac{1}{\left[i(\omega_{n^\prime}-\omega_m)-\xi_\bk\right] \left[i(\omega_{n}-\omega_m)-\xi_\bk\right]}.
\end{eqnarray*}

Since we are interested in the behavior of electrons near the Fermi level, we set 
${\bf k} = {\bf k}_\mathrm{F}$ ($\xi_\bk = 0$). The summation over the internal 
Matsubara frequencies can now be performed exactly, yielding  
\begin{eqnarray*}
\fl\Gamma_{\bk_\mathrm{F}}^{(1)}(\omega_n,\omega_n^\prime)=&\\
  \lambda_\mathrm{m}\left[ \coth\left(\frac{\Omega}{2T}\right)\frac{\Omega^2(\Omega^2-\omega_n\omega_{n^\prime})}
        {(\Omega^2 + \omega^2_n)(\Omega^2+\omega^2_{n^\prime})} 
    - \delta_{n,n^\prime}\frac{\Omega^3/2T}{\Omega^2 + \omega^2_n}\right]. 
\end{eqnarray*}
Here, we have reintroduced the dimensionless coupling 
$\lambda_\mathrm{m} = \frac{g^2_0}{\Omega^2}$ \cite{RademakerPreprint}. 

For inelastic scattering $\omega_n \ne \omega_{n^\prime}$, the vertex is bounded by 
\begin{equation*}
0 < \Gamma^{(1)}_{\bk_\mathrm{F}}(\omega_n,\omega_{n^\prime}) \le \lambda_\mathrm{m}
\coth\left(\frac{\Omega}{2T}\right).
\end{equation*}
Since $\coth(\Omega/2T) \rightarrow 1$ in the relevant regime $T\ll\Omega$, we conclude that 
the vertex correction is bounded by the dimensionless coupling strength $\lambda_\mathrm{m}$. 
Note that this expression does not include the ratio $\Omega/E_\mathrm{F}$ as one would naively 
expect from Migdal's theorem. 

For elastic scattering $\omega_n = \omega_{n^\prime}$, the first order correction diverges 
at low temperatures. 
(Here we are refering to the vertex function for elastic scattering from phonons, 
which is different from the self-energy effects created by elastic forward scattering 
scattering from off-plane impurities that has been discussed in the context of the cuprates \cite{Imp1,Imp2}.) 
This is particularly troublesome, as it leads to a $1/\omega_n$ contribution to the electronic self-energy 
when one includes the first crossing diagram (Fig. \ref{Fig:Diagrams}b). (See also the appendix.) 
Fortunately, this divergence also occurs at higher orders in the self-energy expansion 
and in the limit of perfect forward scattering we sum the elastic vertex corrections in the ladder 
approximation (Fig. \ref{Fig:Diagrams}d). 
We obtain 
\begin{eqnarray*}
\fl\Gamma_{\bk_\mathrm{F}}(\omega_n)=\Gamma^{(0)}_{\bk_\mathrm{F}} (\omega_n) - g_0^2 T \sum_m D^{(0)} ( \omega_m) \times \\ 
\quad\quad\quad\quad\quad	\left[G^{(0)} (\bk_\mathrm{F},  \omega_{n}-\omega_m)\right]^2
		\Gamma_{\bk_\mathrm{F}} (\omega_n - \omega_m) \\
	= 1 - g_0^2 T \sum_m 
		\frac{2 \Omega}{\Omega^2 + \omega_m^2}
		\frac{1}{(\omega_n - \omega_m)^2}
		\Gamma_{\bk_\mathrm{F}} (\omega_n - \omega_m).
\end{eqnarray*}

By performing the Matsubara summation we observe that the dominant term in the low-temperature 
expansion originates from the residue of the double pole at $z = i \omega_n$. We find
\begin{equation}
	\Gamma_{\bk_\mathrm{F}}(\omega_n) =
		1- \lambda_\mathrm{m} \frac{\Omega}{2T} \frac{\Omega^2}{\Omega^2 + \omega_n^2} 
               \Gamma_{\bk_\mathrm{F}}(0).
\end{equation}
It only remains to fix $\Gamma(\bk_\mathrm{F}, 0)$, and so the full elastic vertex function is given by
\begin{equation}
	\Gamma_{\bk_\mathrm{F}}(\omega_n)
		= 1 - \frac{ \lambda_\mathrm{m}}{ \frac{2T}{\Omega} + \lambda_\mathrm{m}} \; \frac{\Omega^2}{\Omega^2 + \omega_n^2}.
\end{equation}
This final result shows that the divergence has disappeared in the summation of the diagrams, 
and the elastic vertex function is bounded between $0 \leq \Gamma \leq 1$.

We conclude that for strong forward scattering, we need to make a distinction between
elastic and inelastic scattering. For inelastic scattering, the vertex
corrections are bounded by the dimensionless coupling $\lambda_m$, which is
already the case for the first order correction. For elastic scattering, we
need to sum up to infinite order in perturbation theory. The resulting elastic
vertex function is bounded by $1$. Furthermore, at low temperatures, the vertex in 
Eq. (7) tends towards zero for $\omega_n$ near the Fermi level. We therefore conclude 
that the elastic contribution to the vertex does not play a strong role in modifying the 
self-energy due to the interaction with the optical phonon. 

\begin{figure}
 \centering
 \includegraphics[width=0.65\columnwidth]{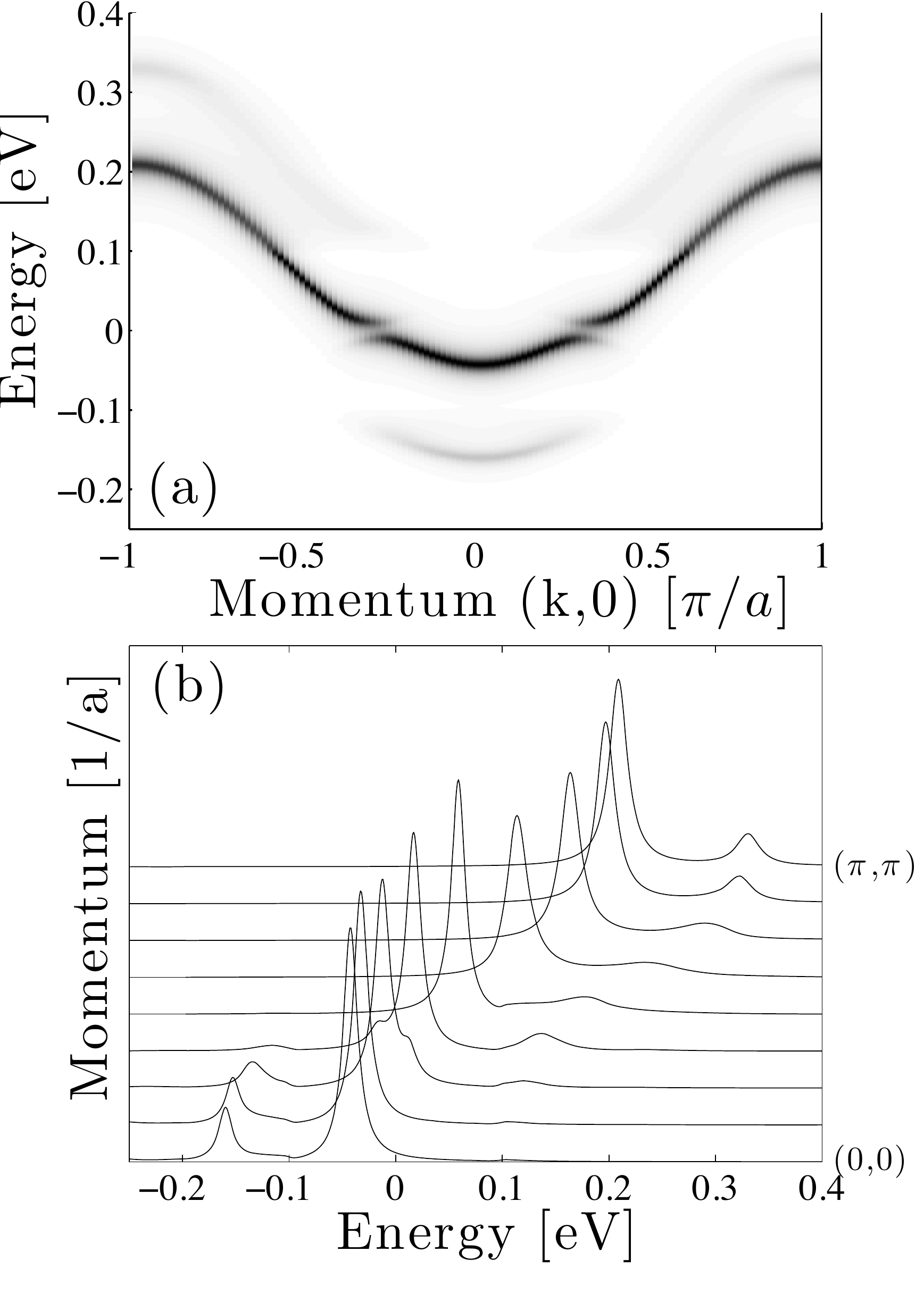}

\caption{\label{Fig:Akw}(a) A false colour image of $A(\bk,\omega)$ at $T = 25$ K for coupling to a 100 meV
optical mode with $q_0 = 0.3/a$ and $\lambda_\mathrm{m} = 0.175$. (b) A waterfall stack of the data shown in (a).}
\end{figure}

There are two important lessons to be drawn from this analysis. First, and most important, the
adiabatic ratio $\Omega/E_\mathrm{F}$ does not appear as a prefactor to
$\Gamma_\bk(\omega_n,\omega_{n^\prime})$. Therefore, the standard measure for validity of Migdal's theorem does
not apply to the case of strong forward scattering, regardless of the location of the band minimum with
respect to the phonon energy. This, however, leads to the second observation. Namely, the vertex correction
is bounded by a term proportional to the dimensionless coupling $\lambda_\mathrm{m}$. Therefore, the
contributions to self-energy due to these corrections will be of order $\lambda^2_\mathrm{m}$ and higher. In
our previous work we showed that $\lambda_\mathrm{m}\sim 0.15\text{--}0.2$ is sufficient to account for the
observed replica bands in the ARPES spectra. As such, the parameter regime relevant to the FeSe/STO system is
in the weak-coupling limit, where the vertex corrections are expected to be small and can be treated
perturbatively. In this context, it is worth noting that this perturbative limit was examined extensively in
Refs. \cite{Vertex1} and \cite{Vertex2}, where it was found that the vertex corrections have a complicated
effect on the superconducting $T_\mathrm{c}$, which depends on the momentum structure of the interaction. In
the perturbative regime, $T_\mathrm{c}$ was found to be suppressed when $g(\bq)$ is dominated by large $\bq$ processes
while it was enhanced when $g(\bq)$ is dominated by small $\bq$. The latter case is the one considered here.

Due to these considerations, we proceed by summing the diagrams traditionally considered by Migdal-Eliashberg
theory and neglect for the time being the vertex corrections. In this case we are required to sum the rainbow
diagrams in order to obtain the superconducting state; however, we note that the majority of our spectral
features are determined by the first order rainbow diagram, which supports the notion that the \ep spectral
features can be captured in the perturbative regime.

\section{Results and Discussion}\label{Sec:Results}
\begin{figure*}
 \centering
 \includegraphics[width=0.9\textwidth]{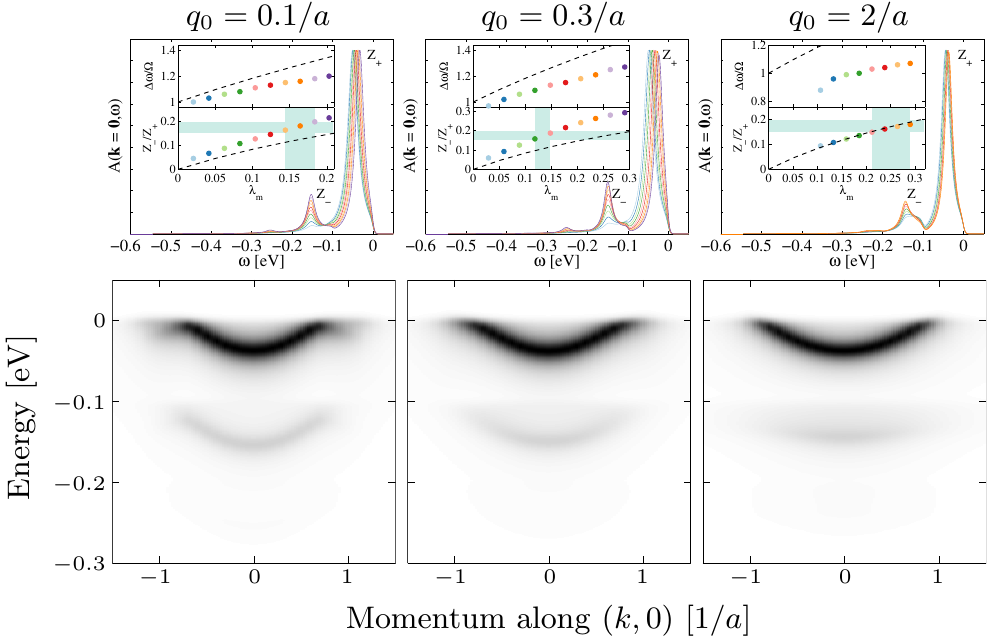}

 \caption{\label{Fig:Akwlmd}Top row: spectral function for a momentum at the
band bottom ($\bk = 0$ in our model, the $M$ point in the experiment) for $T = 30$~K, $q_0 = 0.1/a$ (left),
$0.3/a$ (middle), and $2/a$ (right) and a series of different $\lambda_\mathrm{m}$. $\Omega=100$~meV. The key
feature of the forward scattering mechanism is the appearance of the mirror band ($Z_-$) next to the main
band ($Z_+$). The relative separation $\Delta\omega$ and intensity $Z_-/Z_+ = A(0,\omega_-)/A(0,\omega_+)$ of
these two features is shown in the inset, and increases approximately linearly with $\lambda_\mathrm{m}$ when
$q_0$ is small. The dashed lines show the corresponding result in the perfect forward scattering limit and
the shaded area represents the values of $\lambda_\mathrm{m}$ that are relevant for FeSe/STO. Bottom row:
spectral density $A(\bk,\omega)$ along the $\bk = (k/a,0)$ cut for the corresponding $q_0$ at $T = 30$ K. The
$\lambda_\mathrm{m} = 0.14$, $0.125$, and $0.25$ in the left, middle, and right panels, respectively, which
is a proper value that gives the experimentally observed spectral intensity ratio $Z_-/Z_+ =
0.15\text{--}0.2$ for the corresponding $q_0$ (as indicated by the shaded area in the top panels).}
\end{figure*}

\subsection{The Single-Particle Spectral Function}\label{Sec:Akw}

We begin by presenting details of the spectral properties of our model, which illustrates the formation of
the replica bands. This section will also serve to briefly review some of our results from Ref.
\cite{RademakerPreprint}. Fig.~\ref{Fig:Akw} shows $A(\bk,\omega)$ for $q = 0.3/a$ and $\lambda_\mathrm{m} =
0.175$.  Results are shown for $T = 25$~K $< T_\mathrm{c} = 75.2$~K, and the superconducting gap is apparent
in the main band crossing $E_\mathrm{F}$. Two replica bands are also present, with each offset by
approximately $\Omega$ and $-\Omega$, respectively, from the main band. Here, the forward-focused \ep
interaction produces replicas that are complete copies of the main band; each replica terminates at
$\bk_\mathrm{F}$, and has the characteristic back bending associated with the Bogliubov quasiparticle
dispersion.

The results shown in Fig. \ref{Fig:Akw} are consistent with ARPES data
of Lee {\em et al.} (Ref. \cite{LeeNature2014}). This points to the
$\bq = 0$ nature of the coupling in FeSe/STO \cite{LeeNature2014,RademakerPreprint}.
If the \ep interaction is broadened (increasing $q_0$) then the replica band is
significantly smeared in momentum space, as shown in the lower row of
Fig. \ref{Fig:Akwlmd}. For $q_0 \le 0.3/a$ the replica nature of the
shake-off band is maintained. For $q_0 = 2/a$, however,
the shake-off band is significantly broadened and no longer follows the same
curvature of the main band.
This confirms that $q_0$ must be small in order to
reproduce the experimentally observed clear and complete copies of the main 
band.

The total strength of the interaction can also be extracted from the
data by examining the relative spectral weight of the two bands.
We define the ratio $Z_-/Z_+$ as the ratio of the maximum intensity
in the replica band relative to the main band. The top row
of Fig.~\ref{Fig:Akwlmd} shows the spectral function at $A(\bk=0,\omega)$
for several values of the dimensionless coupling, which allows us to
extract this ratio from our data as a function of $\lambda_\mathrm{m}$. The results
are summarized in the insets of that figure.  A conservative estimate of
$Z_-/Z_+ = 0.15\text{--}0.2$ was obtained from the experimental data in Ref. \cite{LeeNature2014}.
We can compare this to our results, where we find extract $\lambda_\mathrm{m}=0.15\text{--}0.2$
for $q_0 = 0.1/a$ and $\lambda_\mathrm{m}=0.12\text{--}0.15$ for $q_0 = 0.3/a$.
These values confirm that the cross-interface coupling in FeSe/STO is
in the perturbative regime.

\begin{figure*}
 \centering
 \includegraphics[width=0.9\textwidth]{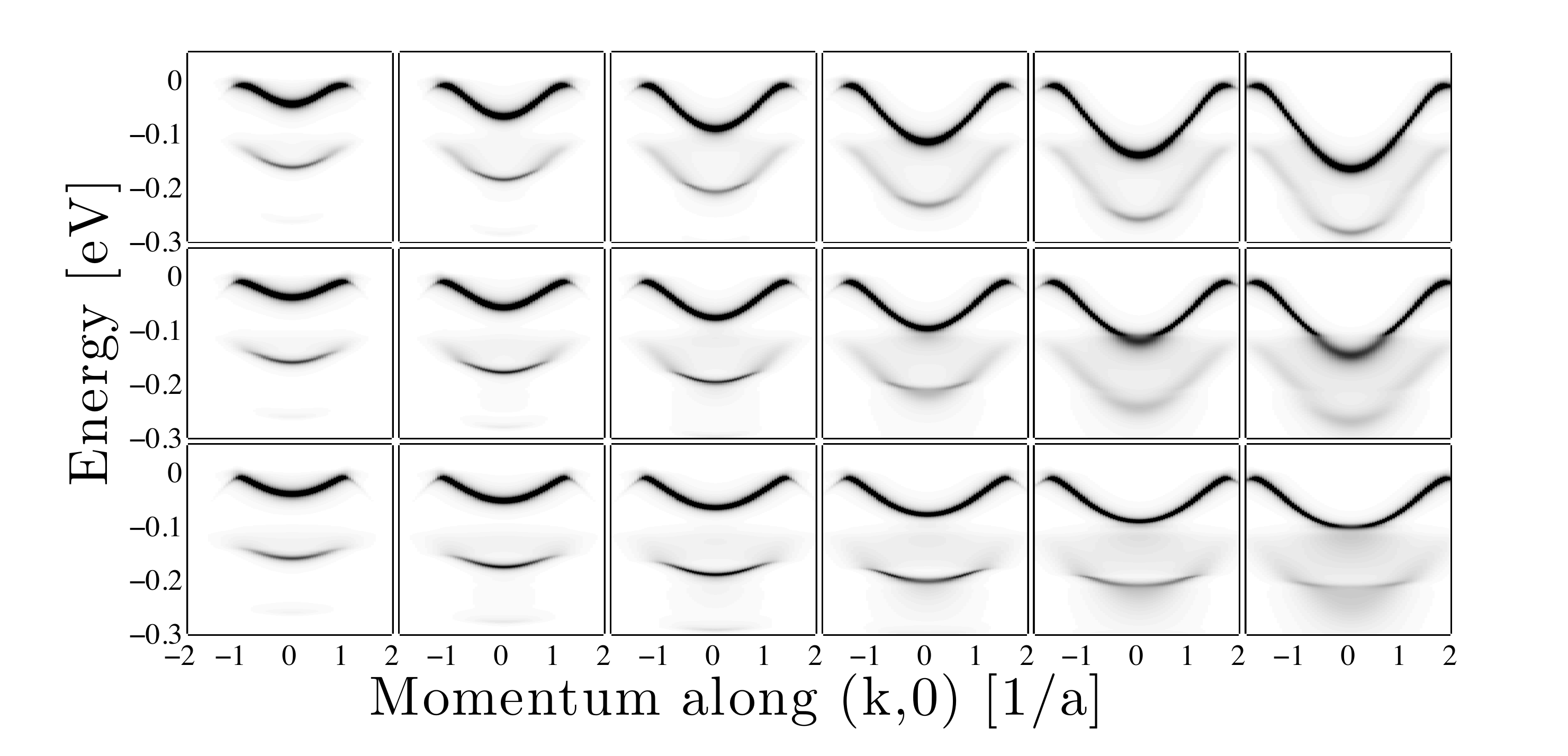}
 \caption{\label{Fig:BandMin}The evolution of the spectral function at $T =
 30$~K as the band minimum crosses the energy scale of the phonon mode $\Omega =
 100$ meV. The top, middle, and bottom rows show results for $q_0 = 0.1/a$,
 $0.5/a$, $2/a$, respectively.  The total strength of the coupling has been
 adjusted to produce $\lambda_\mathrm{m} = 0.2$ in each case.}
\end{figure*}

The interface effects leading to the strong forward scattering peak in the
\ep interaction do not require a shallow electron band as realized in FeSe/STO.
It is therefore interesting to examine how this interaction will manifest
itself in a system where the main band crosses the phonon energy.
In Fig.~\ref{Fig:BandMin} we explored such a scenario by adjusting the chemical
potential of our model such that the band minimum moves from above the
phonon energy (left side)
to below the phonon energy (right side). Results are shown for $q_0 = 0.1/a$ (top
row), $q_0 = 0.3/a$ (middle row), and $q_0 = 2/a$ (bottom row).
In all cases, the replica bands remain located below the main band,
however, as the main band begins to cross the phonon energy scale, we
see the formation of the traditional kink features
commonly seen in the high-$T_\mathrm{c}$ cuprates \cite{CukPSS}. Interestingly,
the kink structure of the main band is also copied completely in the replica band
when $q_0$ is sufficiently small.

\subsection{Properties of the Superconducting State}\label{Sec:Gap}
We now exam the superconducting state mediated by the forward-focused \ep interaction.
Our aim here is not to advocate for a purely phonon mediated pairing interaction in FeSe/STO
at this time. Rather, we wish to examine the effects of an \ep interaction with a strong forward
scattering peak in order to understand the properties of the superconducting state formed by such an interaction.
We will discuss how our results fit into
the larger context of the FeSe/STO discussion in section \ref{Sec:Conclusions}.

\begin{figure}
 \centering
 \includegraphics[width=0.8\columnwidth]{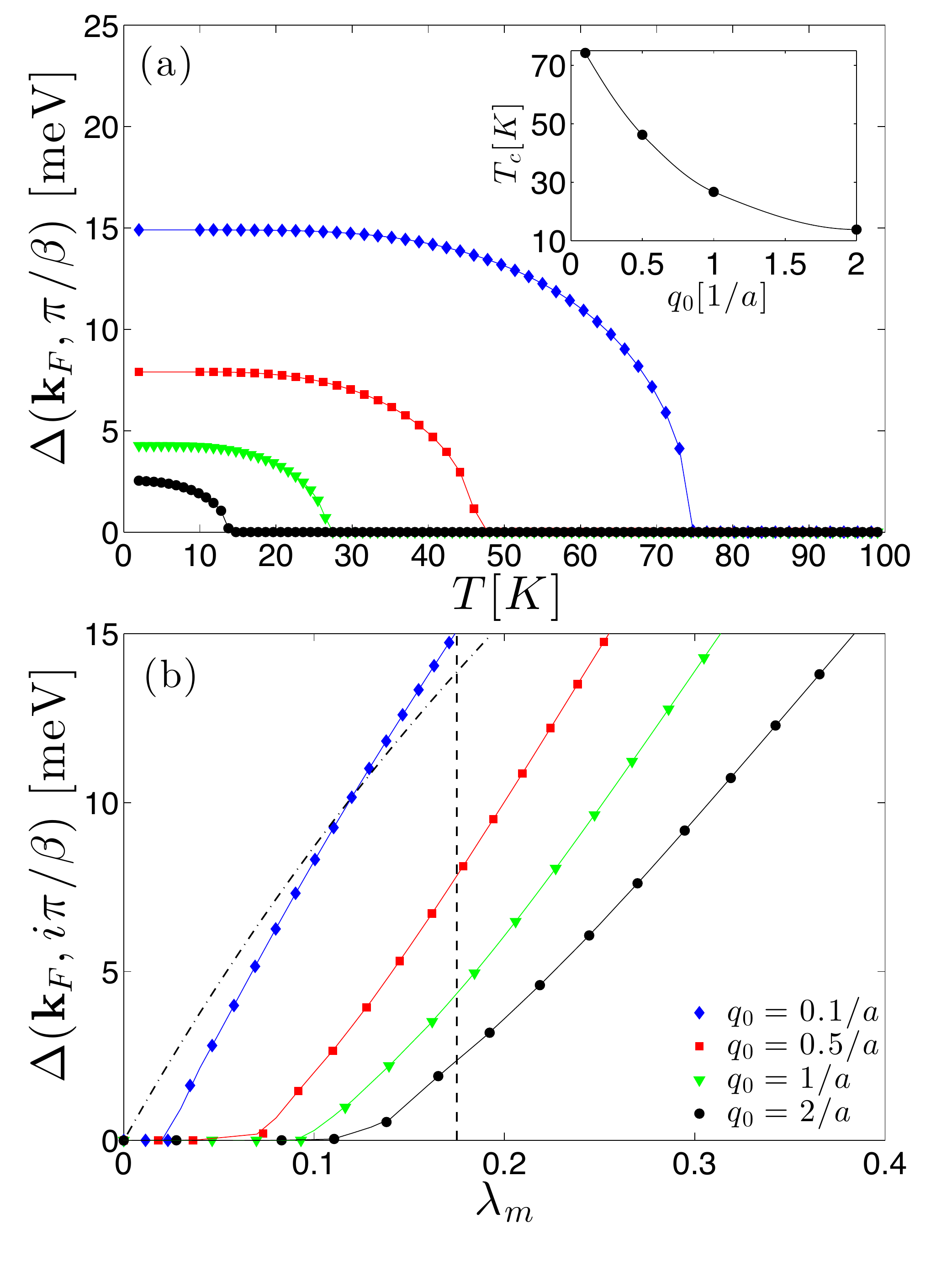}
\caption{\label{Fig:gap}(a) The temperature dependence of the superconducting gap at $\bk_\mathrm{F}$ as a
function of temperature for increasingly forward-focused interactions. Different symbols indicate different
values of $q_0$ used, as indicated by the same legend of (b). (b) The gap value at low temperature ($T=5$~K)
vs. the total coupling strength. The dash-dotted line shows zero temperature gap value expected from the
perfect forward scattering. The dashed line mark the value of the $\lambda_\mathrm{m}$ used in (a).}
\end{figure}

The superconducting gap function at the $n = 1$ Matsubara frequency and on the Fermi surface
$\Delta(\bk_\mathrm{F},i\pi/\beta) = \phi(\bk_\mathrm{F},i\pi/\beta)/Z(\bk_\mathrm{F},i\pi/\beta)$ is plotted
in Fig.~\ref{Fig:gap}(a) as a function of temperature for several values of $q_0$. Only the maximal value on
the Fermi surface is plotted here, since the gap function has $s$-wave symmetry and is quite isotropic for
this set of parameters. Here, results are shown for a series of increasing $q_0$ values, while the total
coupling is fixed at $\lambda_\mathrm{m} = 0.175$. For the broadest interaction ($q_0 = 2/a$) we obtain a
relatively modest $T_\mathrm{c} \sim 13.8$~K, which is larger than the $T_\mathrm{c} \sim 4.3$ K one would
estimate from the weak coupling BCS result. As the interaction becomes more focused around $\bq = 0$,
$T_\mathrm{c}$ increases rapidly, with $T_\mathrm{c}\sim 74.2$ K for $q_0 = 0.1/a$ (see inset). This result
demonstrates the high-$T_\mathrm{c}$'s that can be produced by a strong forward scattering \ep interaction
and are consistent with our prior results \cite{RademakerPreprint}, as well as those obtained from simplified
BCS treatments \cite{DanylenkoEPJB1999}. We obtain $2\Delta/k_\mathrm{B}T_\mathrm{c}$ ratios of $3.76$,
$3.72$, $3.97$, and $4.66$ for $q_0 = 2/a$, $1/a$, $0.5/a$, and $0.1/a$, respectively, indicating that the
deviation from the weak coupling BCS value increases as the interaction becomes increasingly peaked.

The magnitude of the superconducting gap at $T = 5$~K as a function of
$\lambda_\mathrm{m}$ is plotted in Fig.~\ref{Fig:gap}(b) for several values of
$q_0$. For a perfect forward scattering ($q_0\to 0$)
in the weak coupling limit, the analytic expression of the zero temperature
gap on the Fermi surface is
$\Delta_0=2\lambda_m\Omega/(2+3\lambda_m)$ while $T_c=\lambda_m\Omega/(2+3\lambda_m)$ \cite{RademakerPreprint}.
For a strong forward scattering $q_0 = 0.1/a$ we see a linear dependence
on $\lambda_\mathrm{m}$ at weak coupling, in agreement with the result
$T_\mathrm{c} \propto \lambda_\mathrm{m}$ for the same limit.  As the value of $q_0$ is
increased, however, there is a crossover to a more exponential-like dependence
on $\lambda_\mathrm{m}$, which reflects the change from a $T_\mathrm{c} \propto
\lambda_\mathrm{m}$ behavior in the strong forward scattering limit to a
$T_\mathrm{c} \propto \exp(-1/\lambda_\mathrm{m})$ behavior for a more uniform
interaction.

\begin{figure}
 \centering
 \includegraphics[width=0.75\columnwidth]{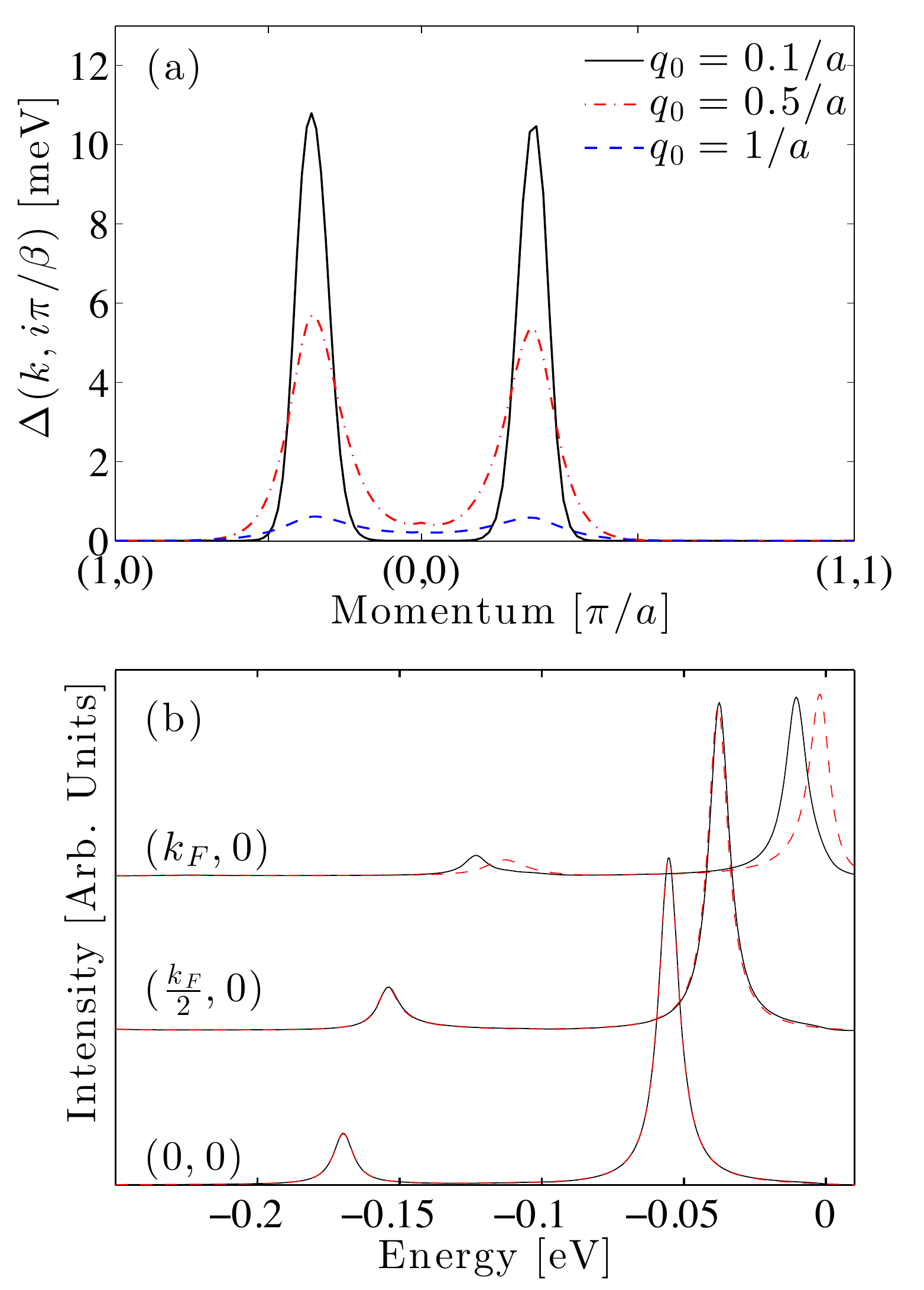}

\caption{\label{Fig:Deltak}(a) The momentum dependence of the superconducting gap function at the first
Matsubara frequency. The results are plotted along the high symmetry directions $(0,0)$ -- $(0,\pi/a)$ (left)
and $(0,0)$ -- $(\pi/a,\pi/a)$ (right). (b) The spectral function obtained for $q_0 = 0.1/a$ at $T = 10$
(solid black) and $100$ K (dashed red). Results are shown for $\bk = (0,0)$,
$\bk = (k_\mathrm{F}/2,0)$, and
$\bk = (k_\mathrm{F},0)$ when read from the bottom to the top of panel (b).}
\end{figure}

Another interesting aspect is the momentum structure of
$\Delta(\bk,i\pi/\beta)$, which is shown in Fig. \ref{Fig:Deltak}. For small $q_0$ the superconducting
gap has maximum values on the Fermi surface but it is
rapidly suppressed for $\bk$ away from $\bk_\mathrm{F}$. This is in contrast to the case of a momentum
independent interaction, where $\Delta(\bk)$ is independent of momentum.
As $q_0$ increases (with $\lambda_\mathrm{m}$ fixed), the total
magnitude of the gap decreases and broadens in momentum space; however,
the maximum remains on the Fermi surface. This behavior can be easily understood in
the limit of perfect forward scattering, where the vertex reduces to $g^2(\bq) = g_0^2N\delta_{\bq,0}$ with
$g_0^2 = \lambda_\mathrm{m}\Omega^2$ \cite{RademakerPreprint}. In the weak coupling limit ($Z = 1$, $\chi = 0$, $\phi
= \Delta$), the gap equation then reduces to
\begin{eqnarray}
\fl  \Delta(\bk,i\omega_n) =& \frac{\lambda_\mathrm{m}\Omega^2}{\beta} \sum_m
    \frac{2\Omega}{\Omega^2 + (\omega_n-\omega_m)^2} \nonumber\\
  &\times \frac{\Delta(\bk,i\omega_m)}{\omega_m^2 + \xi^2_\bk + \Delta^2(\bk,i\omega_m)}.
\end{eqnarray}
This modified gap equation shows that $\Delta(\bk)$ depends most sensitively
on the electronic structure at $\bk$ in the limit of strong forward scattering. Hence the
value of the gap at ${\bf k}$ has decoupled from the values at the other momenta.
As discussed in our previous work \cite{RademakerPreprint}, this ``momentum decoupling" \cite{FS1}
is the ultimate cause of the large $T_\mathrm{c}$ values produced by this type of interaction.
It is also the origin of the momentum structure shown in Fig. \ref{Fig:Deltak}, as one
can immediately see that $\Delta(\bk,i\omega_n)$ will be suppressed as $\xi_\bk$ deviates from the Fermi
surface. This is confirmed by the numerical calculations with finite width $q_0$ of the
interaction shown in Fig.~\ref{Fig:Deltak}(a),
which also show that this effect becomes less pronounced as the width of the interaction
is made more uniform.

The momentum structure of $\Delta(\bk,i\omega_n)$ implies that only the dispersion of states
near $E_\mathrm{F}$ will be affected by the transition to the superconducting state.
This is illustrated in Fig.~\ref{Fig:Deltak}(b), where we show the spectral function
$A(\bk,\omega) = -\mathrm{Im}\hat{G}_\mathrm{11}(\bk,\omega)/\pi$
at $\bk = (0,0)$, $\bk = (k_\mathrm{F}/2,0)$, and $\bk = (k_\mathrm{F},0)$ in the
normal ($T = 100$~K, dashed red) and superconducting ($T = 10$~K, solid black) states.
The peaks of the main and replica bands at $\bk_\mathrm{F}$
shift with the opening of the superconducting gap.
In contrast, the two spectral functions away from $\bk_\mathrm{F}$ are hardly affected in the superconducting
state, since the gap function decays to zero very fast away from the Fermi surface.
Thus, the lack of shift at the band minimum is a characteristic of a superconducting state
mediated by forward scattering bosons. The absence of this shift at the band minimum may be discernible
from the case of a uniform gap if the band is shallow relative to the
size of the gap [$\xi(\bk = 0)\sim \Delta(\bk=0)$].
This condition would be met in FeSe/STO, if the observed $\sim 10 - 15$ meV gap
\cite{LeeNature2014,WangCPL2012,LiuNatureComm2012,TanNatureMaterials2013,HeNatureMaterials2013}
is momentum independent throughout the whole first Brillioun zone.

\begin{figure}[tl]
  \centering
  \includegraphics[width=0.7\columnwidth,draft=false]{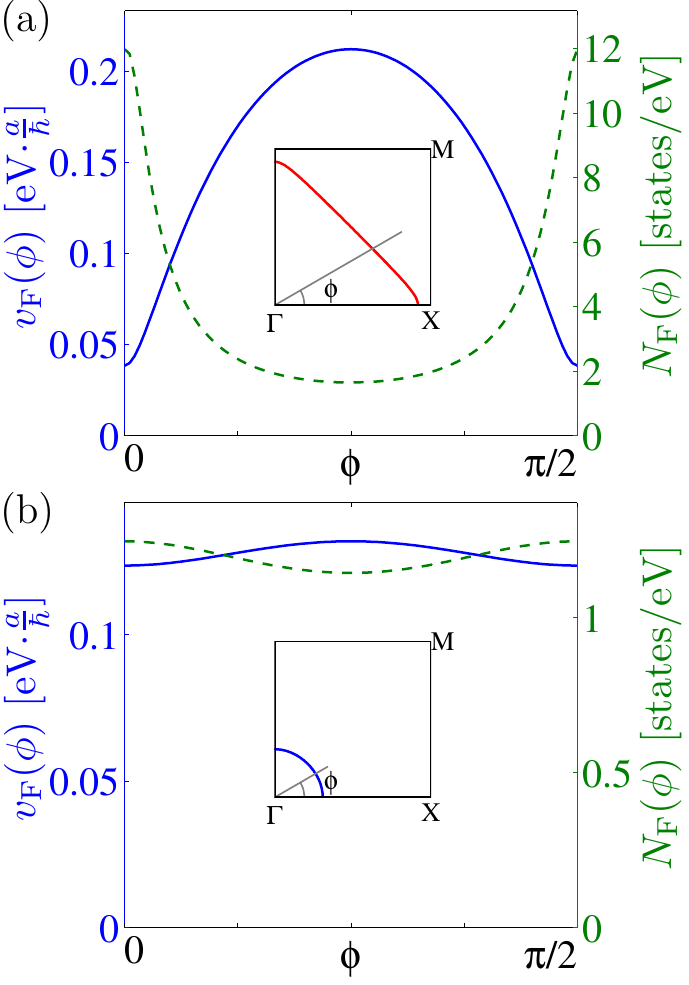}
  \caption{The angular dependence of the $v_\text{F}(\phi)$ (solid line) and local DOS $N(\phi)$ (dashed line)
  for chemical potential (a) $\mu=-5$~meV and (b) $\mu=-235$~meV. The insets
  show the corresponding Fermi surfaces and the definition of the angle around
  the Fermi surface.}
  \label{fig:vFNdos}
\end{figure}

\begin{figure}[tr]
 \centering
  \includegraphics[width=0.7\columnwidth,draft=false]{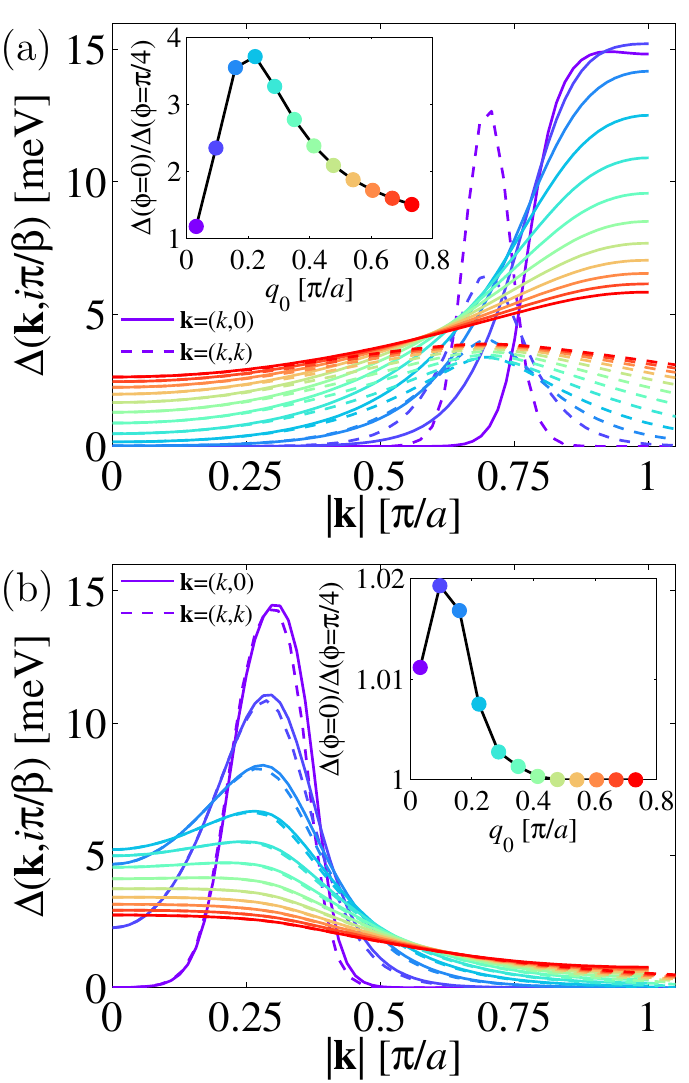}

\caption{Gap function $\Delta(\bk,{\rm i}\pi/\beta)$ at $T=5$~K along the path in the $(k,0)$ (solid line) and
$(k,k)$ directions (dashed line) for chemical potentials (a) $\mu=-5$~meV and (b) $\mu=-235$~meV. The insets
show the ratio of the maximal gap in the $(k,0)$ and $(k,k)$ directions. The colour of lines indicates
different values of $q_0$ as indicated the insets. In all cases $\lambda_\mathrm{m} = 0.2$.}
 \label{fig:gap0kpath}
\end{figure}

Another consequence of the momentum decoupling in the gap equation is that a significant anisotropy in
$\Delta(\bk_\mathrm{F})$ can occur, which is determined by the anisotropy of the Fermi velocity or local DOS
in $\bk$-space~\cite{FS1}. To illustrate this point, we calculated the gap function for our model with $\mu =
-5$~meV such that $E_\mathrm{F}$ lies in close proximity to the van Hove singularity in the band structure.
In this case, the local DOS at Fermi level, defined as $N(\phi)\equiv \frac{1}{2\pi v_\text{F}(\phi)}
\frac{{\rm d}k_{\parallel}}{{\rm d}\phi}$, is extremely anisotropic, as shown in Fig. \ref{fig:vFNdos}(a).
($\phi$ defines the angle around the Fermi surface measured from the positive $x$-axis and $k_{\parallel}$ is
the momentum in the direction tangential to the Fermi surface.) For comparison, $N(\phi)$ and
$v_\text{F}(\phi)$ for $\mu=-235$~meV are shown in Fig. \ref{fig:vFNdos}(b), where each has a much weaker
variation around the Fermi surface. The corresponding gap functions $\Delta(\bk,{\rm i}\pi/\beta)$ at $T =
5$~K along the $(k,0)$ (solid lines) and $(k,k)$ (dashed lines) directions are plotted in
Fig.~\ref{fig:gap0kpath}. Results are shown for several values of $q_0$, as indicated by the colour of the
lines. The insets of Fig.~\ref{fig:gap0kpath} show the ratio of the gap maximum along the $(k,0)$ ($\phi =
0$) and $(k,k)$ ($\phi = 45^\circ$) directions, which defines a measure of the gap anisotropy.

In the $\mu=-5$~meV case, we see a non-trivial evolution of the gap anisotropy as a function of $q_0$. For
large $q_0$ (red), i.e.~closer to uniform scattering, $\Delta(\bk,i\omega_n)$ is somewhat anisotropic with
the maximum value along ($k,0$) being $\sim 1.5$ times the maximum along the diagonal. These values are
inversely proportional to the Fermi velocity, with $\Delta \propto N(\phi) \propto 1/v_\mathrm{F}(\phi)$,
which can be inferred from the gap equation in the perfect forward scattering limit. As the value of $q_0$
initially decreases, the gap anisotropy becomes more pronounced until a maximum is reached for $q_0 \sim
0.6/a$. This reflects the momentum decoupling effect seen in Ref.~\cite{FS1}. Further decreases in $q_0$,
however, result in a more uniform gap. This can be traced to an increase in the gap magnitude along the zone
diagonal relative to the gap along $(k,0)$. We speculate that the decreasing anisotropy in this region is
related to an increase in the phase stiffness of the condensate when $q_0$ is very small and larger total gap
values are obtained. These results are in contrast to those obtained when $\mu = -235$~meV and the anisotropy
in the band structure is much less pronounced. In this case, the resulting gap function is relatively
isotropic with at most a $\sim 2\%$ variation occurring when $q_0 = 0.3/a$. In both cases the gap has an
$s$-wave symmetry, as can be expected for an attractive interaction.

Figs.~\ref{fig:vFNdos} and \ref{fig:gap0kpath} demonstrate that a strong forward scattering can produce an
anisotropic gap if the variations in the Fermi velocity are significant. In such a case the gap magnitude is
inversely related to the Fermi velocity. Early ARPES measurements on FeSe/STO found that the superconducting
gap was fairly isotropic \cite{LiuNatureComm2012,TanNatureMaterials2013}; however, a more recent ARPES study
has observed a $\sim 30\%$ variation in $\Delta(\bk)$ around the electron pockets \cite{GapAnisotropy}. This
is larger than we infer from our single-band model, but our calculation neglects the nearly degenerate
electron pockets, their relative orbital character, and any hybridization effects between them, and the
anisotropy from an unconventional pairing mechanism that may be working in conjunction with the forward
scattering. It would be interesting to carry out our calculation for a more realistic band structure that
captures these aspects.

The last topic that we wish to discuss in this section is the relevance for forward
scattering to pairing channels that are
traditionally considered in the context of purely electronic, repulsive interactions.
For any momentum dependent interaction, one can define a projection of the coupling
in the $\alpha$-pairing channel \cite{JohnstonPRB2010}
\begin{equation}
\lambda_\alpha = \frac{2}{N\Omega}\frac{\sum_{\bk,\bp}|g(\bk-\bp)|^2Y_\alpha(\bk)Y_\alpha(\bp)\delta(\xi_\bk)\delta(\xi_\bp)}
{\sum_{\bk} Y^2_\alpha(\bk)\delta(\xi_\bk)},
\end{equation}
where $Y_{s} = 1$ for an $s$-wave gap, $Y_{\pm s} = \cos(k_xa)\cos(k_ya)$ for an $s_\pm$-wave gap,
$Y_{d_{x^2-y^2}} = [\cos(k_xa)-\cos(k_ya)]/{2}$ for a $d_{x^2-y^2}$ gap, and so forth. If
$\lambda_\alpha > 0$ then the coupling is attractive in that channel. For a single-band system with strong
forward scattering one can convince oneself that $\lambda_{\alpha}$ is positive for many $\alpha$ since 
$Y_\alpha(\bk)$ and $Y_\alpha(\bp)$ have a better chance to have the same sign if $\bk$ is close to $\bp$
when $|g(\bk-\bp)|$ is large (this is somewhat sensitive to shape and size of the Fermi surface) while
$\lambda_s$ is the largest. As a result, strong forward scattering on its own will realize an $s$-wave gap but has
non-zero contributions in other pairing channels. Therefore, if a dominant pairing interaction arising from
repulsive interactions is present, then the forward scattering \ep interaction can contribute to the
resulting unconventional superconducting state in most instances.

\begin{figure}[h]
 \centering
 \includegraphics[width=0.9\columnwidth]{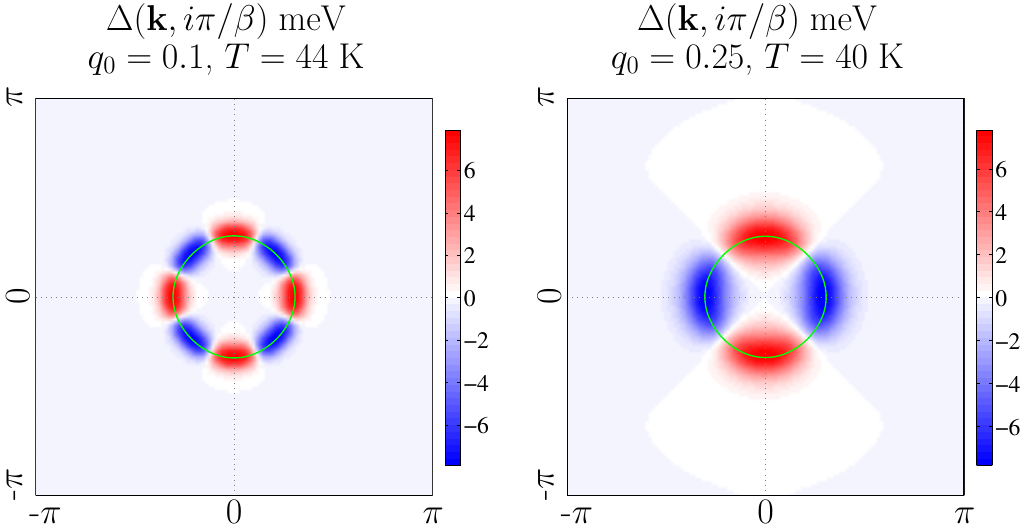}

\caption{\label{Fig:gapfwvdwv}Left: $f$-wave gap $\Delta(\bk,i\omega_{n=1})$ for $q_0=0.1$. Right:
$d$-wave gap $\Delta(\bk,i\omega_{n=1})$ for $q_0=0.25$. $\mu^*=0.2$, $\lambda_\mathrm{m}=0.14$ for $q_0=0.1$,
and $\lambda_\mathrm{m}=0.18$ for $q_0=0.25$ are used in the calculation. The green line indicates the Fermi surface.}
\end{figure}

These considerations also show that if the $s$-wave channel is blocked by a repulsive interaction in
a singleband system, then the
forward scattering will produce a gap symmetry in the channel with the next leading value
of $\lambda_\alpha$. This was pointed out
in Ref.~\cite{Santi1996}, where it was discussed in the context of forward scattering in the cuprates using
a simplified Eliashberg treatment. We demonstrate this here for our model by introducing
the Morel-Anderson Coulomb pseudopotential $\mu^{*}$, which suppresses the $s$-wave pairing
tendency. Fig. \ref{Fig:gapfwvdwv} shows the
resulting gap function for two different choices in parameters and $\mu^* = 0.2$.
In this case the $s$-wave channel is effectively blocked by $\mu^*$ and a $d$-wave or
$f$-wave gap is realized, depending on our parameter choice. We stress that these solutions
were not imposed during our self-consistent calculations. Rather, they were found naturally
during the iterative procedure.

The reader should also note that the $T_\mathrm{c}$ is not strongly suppressed by the Coulomb
pseudopotential~\cite{RademakerPreprint}.

\subsection{The Single-particle Density of States}\label{Sec:DOS}
\begin{figure}
  \centering
  \includegraphics[width=0.45\textwidth]{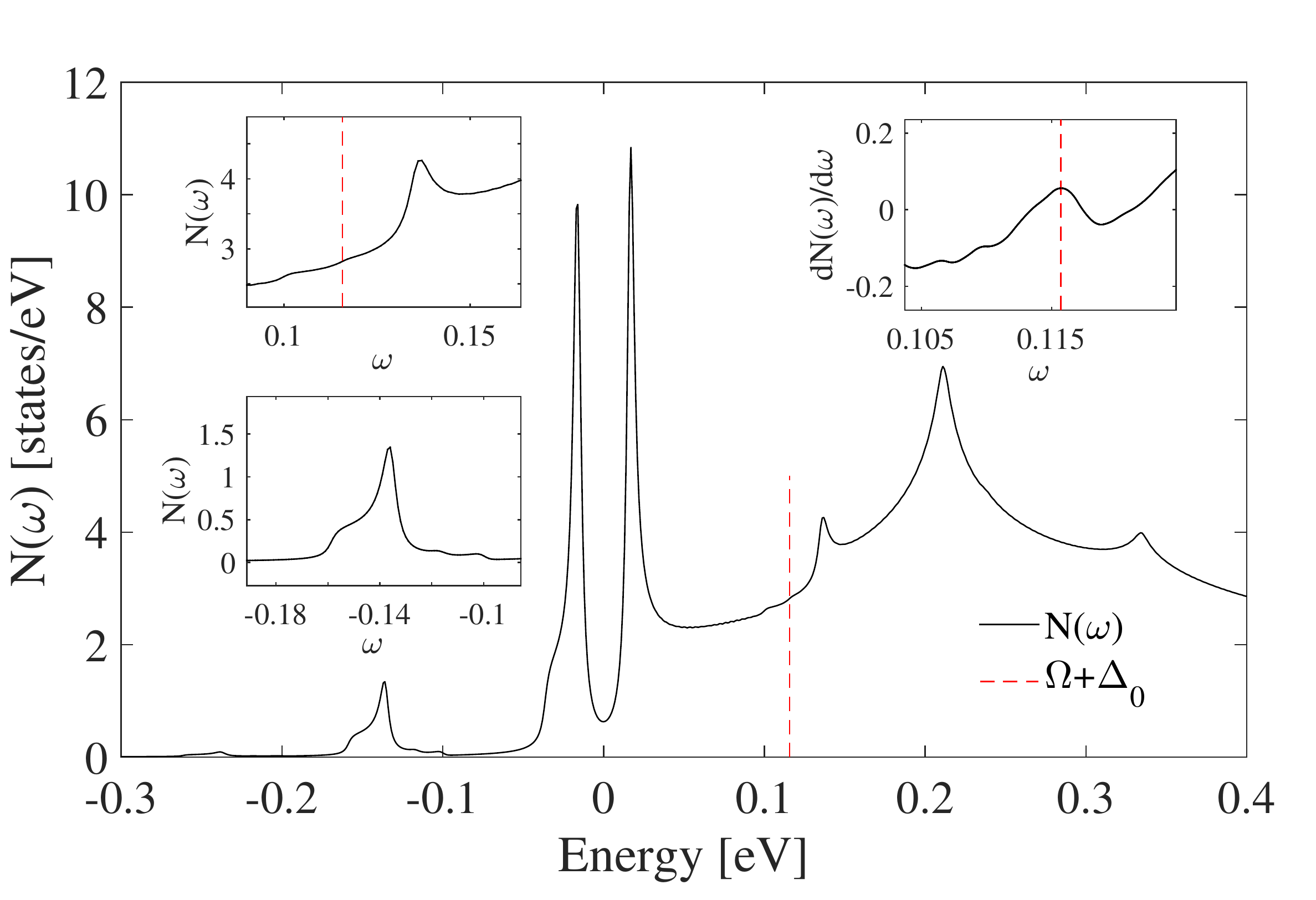}

\caption{\label{Fig:DOS}The density of states $N(\omega)$ for $q_0 = 0.1/a$, $\lambda_\mathrm{m} = 0.19$, and
$\Omega = 100$ meV, calculated at $T = 10$ K. A small $\gamma = 3$ meV contribution was also added to the
imaginary part of the single-particle self-energy in order to broaden the spectra at this $\bk$-resolution.
The two insets on the left zoom in on the spectra in the regions where the boson-induced features are found and
the inset on the right shows ${\rm d}N(\omega)/{\rm d}\omega$.}
\end{figure}

In a boson-mediated superconductor, the electron-boson interaction produces fine-structure in the 
single-particle
density of states (DOS) $N(\omega)$. The energy at which these structures appear is determined by boson
energy shifted by the maximum value of the superconducting gap $\Omega + \Delta_0$ \cite{McMillanParks}. In
addition, these structures can appear either as shoulders or as local minima in $N(\omega)$, depending on the
gap symmetry and momentum dependence of the interaction \cite{JohnstonPRB2010,deCastroPRL2008}. As such, the
renormalized electronic structure carries valuable information regarding the pairing mediator. Motivated by
this, we calculated $N(\omega) = -\frac{2}{\pi N} \sum_\bk \mathrm{Im}\hat{G}_{11}(\bk,\omega)$ for our model
in the case of a strongly forward focused interaction. The result is shown in Fig.~\ref{Fig:DOS}, where the
model parameters are $\Omega = 100$~meV, $q_0 = 0.1/a$, and $\lambda_\mathrm{m} = 0.19$. These results were
obtained on an $256\times 256$ $k$-grid and at a temperature of $T = 10$~K, well below $T_\mathrm{c}$.

The DOS shows clear coherence peaks associated with the fully gapped $s$-wave
state, which are broadened by a $3$~meV contribution that was added
to the imaginary part of $Z(\bk,\omega)$. The DOS also has several additional
structures arising from the replica bands. The insets of Fig.~\ref{Fig:DOS}
zoom in on the relevant energy ranges. In this case, the structures are
different from those expected for a conventional \ep mediated superconductor
with an $s$-wave gap. In the conventional case, the boson renormalizations form
shoulders in $N(\omega)$, where the DOS is enhanced (suppressed) for energies
below (above) $\Omega+\Delta_0$, resulting in a characteristic minima in
$dN/d\omega$ \cite{McMillanParks}. In the forward scattering case, however, we
see step-like features located at approximately $\Omega$ above and below
$E_\mathrm{F}$, which correspond to the onset of the spectral weight of the
replica bands.  This results in a local maxima in $dN/d\omega$ at the energy
$\Omega$, occurring for both for the unoccupied and occupied DOS. This
difference provides another structure in the electronic structure that might be
used for identifying replica bands in real materials.

\subsection{Quasiparticle Interference}\label{Sec:QPI}
Fourier transform scanning tunneling spectroscopy, also commonly known as
quasi-particle interference (QPI), has become a popular method for studying the
electronic structure and gap symmetry of unconventional superconductors
\cite{QPI1,QPI2,QPI3,QPI4,QPI5,QPI6,FanPreprint}. This technique exploits the
fact that the incoming and outgoing wavefunction of an electron scattered from
an impurity will interfere with each other to form ripples in the charge
density. The power spectrum of the resulting density modulations measured at a
bias voltage $e\mathrm{V} = \omega$ will have peaks in its intensity at wave vectors
$\bq(\omega)$, which are related to spanning vectors across the constant energy
contours of the band $\xi(\bk) = \omega$. By tracking the dispersion of the
$\bq(\omega)$ peaks, one can infer information about the underlying electronic
band structure. One of the advantages of this technique over ARPES is that it
can access unoccupied states \cite{HuangPRL2015}.  This technique can therefore
probe both the occupied and unoccupied replica bands produced by a strong
forward scattering mode. Motivated by this, we calculated the QPI spectrum for
our model.

\begin{figure*}
  \centering
  \includegraphics[width=\textwidth,trim=0in 2in 0.3in 0in, clip=true]{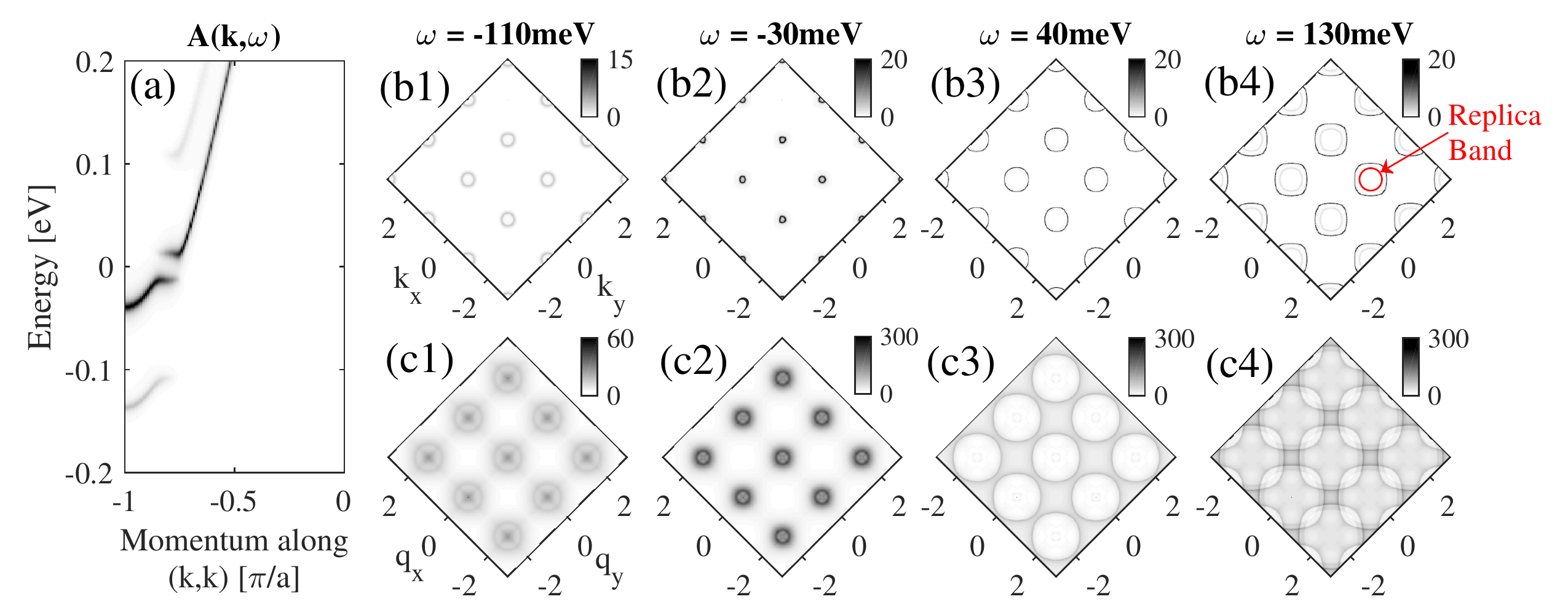}

\caption{\label{Fig:QPI}A summary of the electronic structure and quasiparticle interference (QPI) spectrum
for our model. Results are shown for the case of a strong forward scattering $q_0 = 0.1/a$,
$\lambda_\mathrm{m} = 0.175$, $\Omega = 100$~meV, and at a temperature $T = 10$~K $\ll T_\mathrm{c}$. (a) The
single-particle spectral function along the zone diagonal. (b1)--(b4) $A(\bk,\omega)$ for fixed values of
$\omega$ as indicated. (c1)--(c4) The corresponding QPI intensities $|\delta\rho(\bq,\omega)|$ as a function
of $\bq$. In (b4) one of the inner ring with relatively weak intensity is marked by a red circle.}
\end{figure*}

Our QPI calculations are carried out using the standard $T$-matrix formalism
for a single-band superconductor, as outlined in a number of references
\cite{WangPRB2003,CapriottiPRB2003,GrothePRL2013}. In the presence of an
impurity, the Fourier transform (FT) of the local electron density (FT-LDOS) is
partitioned as $\rho(\bq,\omega) = \rho_0(\bq,\omega) +
\delta\rho(\bq,\omega)$, where $\rho_0(\bq,\omega)$ is the FT-LDOS for the
homogeneous system and $\delta\rho(\bq,\omega)$ is the FT of the induced
density modulations due to the impurity. The latter part is given by
$\delta \rho(\bq,\omega) = \frac{i}{2\pi} g(\bq,\omega)$, where
\begin{eqnarray}
\fl  g(\bq,\omega) =& \frac{1}{N}\sum_\bk\Big[\hat{G}_{11}(\bk,\bq,\omega) + \hat{G}_{22}(\bk,\bq,-\omega) \nonumber\\
  &  - \hat{G}^*_{11}(\bk,-\bq,\omega) - \hat{G}^*_{22}(\bk,-\bq,-\omega) \Big].
\end{eqnarray}
Here, $\hat{G}(\bk,\bq,\omega) =
\hat{G}_0(\bk+\bq,\omega)\hat{T}(\bk+\bq,\bk,\omega) \hat{G}_0(\bk,\omega)$ is
the Green's function in the presence of the impurity; $\hat{T}(\bp,\bk,\omega)$
is the $T$-matrix; and $\hat{G}_0(\bk,\omega)$ is the bare Green's function in
the absence of the impurity. For our purposes in this section,
$\hat{G}_0(\bk,\omega)$ includes the self-energy due to the \ep interaction.
The assumption here is that the impurities and \ep interaction do not
significantly influence one another.

For simplicity, we treat the impurity as a weak non-magnetic point-like potential scatter
located at the origin. In this limit the $T$-matrix is momentum independent and is given by
\begin{equation}
  \hat{T}(\omega) = \left[\hat{V}^{-1} - \frac{1}{N}\sum_\bk \hat{G}_0(\bk,\omega)\right]^{-1},
\end{equation}
where $\hat{V} = V_0\hat{\tau}_3$ and $V_0$ characterizes the strength of the impurity potential. For our
calculations we took $V_0 = 10$~meV; however, our results are not too sensitive to this value.

\begin{figure*}[t]
  \centering
  \includegraphics[width=0.8\textwidth,trim=0in 0in 0in 0in, clip=true]{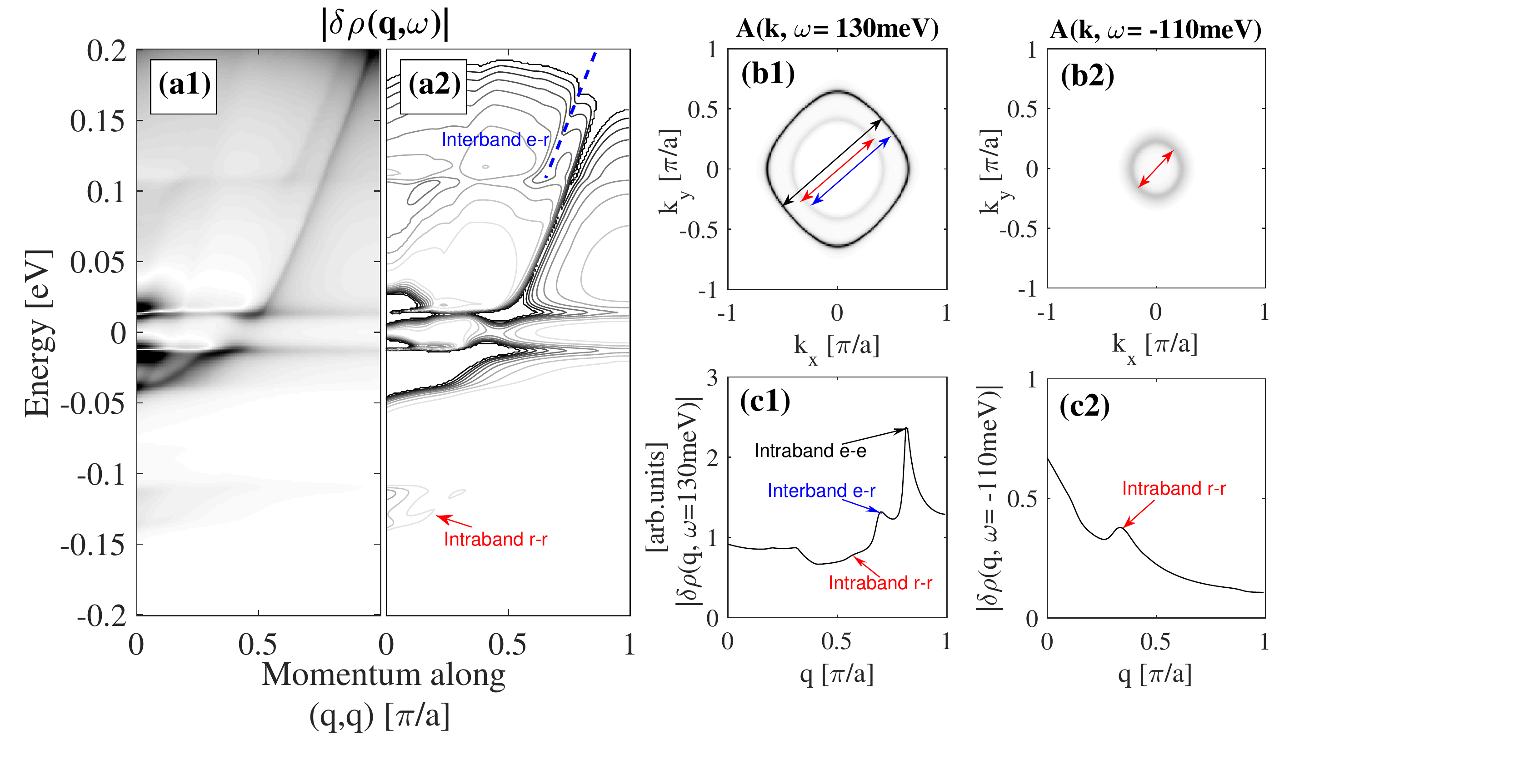}

\caption{\label{Fig:QPI_dispersion}(a1), (a2) The energy-momentum dispersion of the QPI intensity
 as a false colour and contour plot, respectively. The red arrows indicate the QPI feature arising
 from intraband scattering within the occupied portion of the replica band. The blue dashed lines
 indicate the QPI features arising from the scattering between the unoccupied states of the main
 and replica bands. (b1), (b2) constant energy cuts of the spectral function at $130$ and $-110$ meV,
 respectively. The possible scattering processes are indicated by the arrows. (c1), (c2) the QPI
 intensity at $130$ and $-110$~meV, respectively, for a cut along the diagonal direction
 $\bq = (q,q)$. In panels (a2), (c1), and (c2) ``e" denotes the main electron band and ``r" denotes the 
 replica band.
 }
\end{figure*}

Our QPI results are summarized in Figs.~\ref{Fig:QPI} and
\ref{Fig:QPI_dispersion}.  The spectral function for our ``bare" electronic
structure, including the effects of the \ep interaction, is shown in
Fig.~\ref{Fig:QPI}(a) for reference. The relevant model parameters are given in
the figure caption.  Constant energy cuts of the spectral function are shown
Figs.~\ref{Fig:QPI}(b1)--(b4) for energies located inside the occupied replica
band [$-110$~meV, panel (b1)], inside the occupied states of the main band
[$-30$~meV, panel (b2)], inside the unoccupied states of the main band
[$40$~meV, panel (b3)], and inside the unoccupied replica band [$130$~meV,
panel (b4)]. In Fig.~\ref{Fig:QPI}(b1)--(b3) one can see a clear ring-like
feature corresponding to the single replica or main band that is intersected at
these energies; however, in Fig.~\ref{Fig:QPI}(b4), a double ring structure is
observed where the inner ring corresponds to the replica band and the outer
ring corresponds to the main band.

This electronic structure is reflected in the corresponding QPI spectra, shown in
Fig.~\ref{Fig:QPI}(c1)--(c4), respectively. The maps in Figs.~\ref{Fig:QPI}(c1)--(c3) have a clear ring-like
structure with a radius $\bq = 2\bk(\omega)$, where $\bk(\omega)$ is a vector spanning the constant energy
contours. These features resemble what one would expect for a simple single band system 
\cite{GrothePRL2013, CapriottiPRB2003}.  At $\omega =
130$~meV, however, the QPI pattern develops a double ring structure, as one would might expect from the
topology of the main and replica bands shown in Fig.~\ref{Fig:QPI}(b4). Based on the geometry of the features
we can assign the outer QPI peak to intraband scattering within the main band and inner QPI peak to interband
scattering between the unoccupied states of the replica and main bands. We also see a very weak feature due
to intraband scattering within the replica band, however, it is very difficult to resolve in this image. The
intensity of these features also follows the hierarchy of spectral weight associated with the type of band.
The relative intensity of these features are also shown in Fig.~\ref{Fig:QPI_dispersion}.

These results establish that the replica bands can be observed in the QPI
spectra and therefore FT-STS could in principle study these features,
particularly on the unoccupied side of $E_\mathrm{F}$.  In this case, the QPI
peaks due to the replica bands will appear similar to regular bands at a given
bias voltage albeit with a reduced intensity; however, the dispersion and
spectral weight of these features will be quite different. To illustrate this,
we plot the QPI intensity along the $\bq = (q,q)$ line as a function of energy
in order to track the dispersion of the QPI peaks.
Fig.~\ref{Fig:QPI_dispersion}(a1) presents this information as a false colour
image while Fig.~\ref{Fig:QPI_dispersion}(a2) presents it as a contour plot in
order to highlight the features with weaker intensity.  If a QPI peak is due to
intraband scattering in the main band, then it will disperse towards $\bq = 0$
as the bias voltage is tuned towards the top or bottom of the band
\cite{GrothePRL2013}. This can be clearly seen in
Fig.~\ref{Fig:QPI_dispersion}(a1) from the most intense feature. In contrast, the
dispersion of a peak due to inter or intraband scattering to a replica band
will be different due to the extinction of the replica band's spectral weight
at $\bk_\mathrm{F}$. This is apparent in Fig.~\ref{Fig:QPI_dispersion}(a1) and
Fig.~\ref{Fig:QPI_dispersion}(a2), where the intensities of the interband
scattering feature at positive energies and the intraband scattering feature at
negative energies sudden disappears at the energy where the replica band
disappears.  This is another unique spectral fingerprint of a replica band.

While there have been several QPI studies conducted on the FeSe/STO system
\cite{FanPreprint,HuangPRL2015}, to the best of our knowledge no indications of the replica bands have
been observed. One study by Huang {\it et al.} (Ref. \cite{HuangPRL2015}) found evidence for an electron-like
and located 75 meV above $E_\mathrm{F}$; however, the measurements of the tunneling decay length indicated
that this band was centered at the $\Gamma$-point rather than the $M$-point.
This rules out the possibility
that this feature is due to the replica expected for the electron pockets at the $M$-point.
However, we also note that nearly all of the QPI studies completed
to date have focused on energy windows of less than 80 meV of the Fermi level
\cite{FanPreprint,HuangPRL2015}. This is outside the region where the replica band are expected, based on the
phonon energy scale inferred by ARPES \cite{LeeNature2014}. It would therefore be interesting to extend these
studies to higher bias voltages in order to determine if the features shown in
Fig.~\ref{Fig:QPI_dispersion} can be observed. This
will be a challenging task, however, as our calculations have shown that the replica QPI peaks 
are much weaker in intensity than the main QPI peaks.

\section{Discussion and Concluding Remarks}\label{Sec:Conclusions}
We have examined several aspects of a system of electrons coupled to an optical phonon mode via a strongly
momentum dependent interaction, which is peaked for small momentum transfers $\bq$. This interaction is
remarkably effective at mediating pairing, resulting in a superconducting state with a high-$T_\mathrm{c}$
for very modest values of the total dimensionless coupling $\lambda_\mathrm{m}$ and a rich phenomenology.
In the absence of a Coulomb pseudopotential the superconducting state has an $s$-wave symmetry.
But the gap function can be anisotropic when there are significant variations in the
system's Fermi velocity and the interaction is sufficiently peaked at $\bq = 0$. In addition, 
other gap symmetries including $d$- and $f$- wave symmetries can be realized
when the $s$-wave channel is blocked by the Coulomb interaction.
This result highlights the property that a
forward scattering interaction can be attractive in other pairing channels that are
traditionally mediated by repulsive interactions.

The forward-focused interaction leaves several distinct replica bands in the electronic structure, in
agreement with those observed in two recent ARPES studies \cite{PengNatureComm2014,LeeNature2014} Here, we
have further demonstrated that these features can also be found in the single-particle DOS and QPI patterns
probed by STM/STS. More importantly, the manifestation of these features in the DOS is qualitatively
different than the usual DOS modulations induced by the conventional attractive \ep interaction mediated by
phonons. Similarly, the intensity and dispersion of the QPI peaks due to scattering from the replica bands
have a distinctive energy dependence. These facts allow the case of forward scattering to be differentiated
from the case of a more conventional (uniform) \ep interaction. These predictions can help future experiments to
confirm or rule out the presence of such an interaction in FeSe/STO, FeSe/BTO, or other novel interface
systems.

We close with some comments on the situation of FeSe films on oxide substrates. The forward scattering
mechanism is very effective at producing high values of $T_\mathrm{c}$. Moreover, the richness of the pairing
symmetries we observed here further confirms that this interaction can also help mediate unconventional
pairing channels. Combined, these facts mean that the cross interface coupling with a strong $\bq = 0$ peak
can be an effective means to engineer superconductivity at higher transition temperatures in low-dimensional
interfaces. As discussed in the introduction, and in Ref. \cite{LeePreprint}, the intercalated FeSe systems
have band structures remarkably similar to FeSe/STO and a $T_\mathrm{c} \sim 40$~K. Similarly, FeSe thin
films post treated with K and Na adatoms result in comparable values of $T_\mathrm{c}$. These factors suggest
that an unconventional pairing mechanism is most likely present; however, the maximal $T_\mathrm{c}$ that
such a pairing mechanism produces appears to be pinned below $40$~K. As argued in Ref. \cite{LeePreprint},
the cross interface coupling can provide the additional increase in superconductivity needed to reach the
observed $T_\mathrm{c} \sim 55-75$~K. There is also the strong possibility that the incipient holelike band
at the $\Gamma$ point contributes to the traditional $s_\pm$ pairing scenario due to the large energy scale
of the spin fluctuations \cite{WangEPL,BangNJP,ChenPreprint}. In this case, the small $\bq$ scattering
introduced by the phonons would be primarily intraband in nature and thus benificial to pairing. Multichannel
scenarios such as these may also shed light on the recent observation of two superconducting domes
\cite{SongPreprint} in K doped-FeSe/STO, a result that highlights the richness of the superconducting state
in this system and the possibility of two pairing mechanisms. As we concluded in Ref.
\cite{RademakerPreprint}, one obvious experiment that would potentially help sort out this situation is an
$^{18}$O isotope measurement for $T_\mathrm{c}$. If a joint pairing mechanism is realized then the $\alpha =
\frac{1}{2}$ isotope coefficient should be reduced by a significant amount \cite{Isotope1,Isotope2}. It
should be noted, however, that vertex corrections to the \ep interaction can also modify the expected isotope
coefficient \cite{Vertex2}.

\ack%
We thank M. Berciu, P. J. Hirschfeld, A. Linscheid, and D. J. Scalapino for useful conversations. L. R.
acknowledges funding from Rubicon (Dutch Science Foundation). A portion of this research was conducted at the
Center for Nanophase Materials Sciences, which is a DOE Office of Science User Facility. 
This manuscript has been authored by UT-Battelle, LLC under Contract No. DE-AC05-00OR22725 with the U.S. 
Department of Energy. The United States Government retains and the publisher, by accepting the article for  
publication, acknowledges that the United States Government retains a non-exclusive, paid-up, irrevocable, world-
wide license to publish or reproduce the published form of this manuscript, or allow others to do so, for United 
States Government purposes. The Department of Energy will provide public access to these results of federally  
sponsored research in accordance with the DOE Public Access Plan (http://energy.gov/downloads/doepublic-
access-plan). CPU time was
provided in part by resources supported by the University of Tennessee and Oak Ridge National Laboratory
Joint Institute for Computational Sciences (http://www.jics.utk.edu).

\appendix
\section*{Appendix}
\setcounter{section}{1}
In the perfect forward scattering limit, 
the contribution to the self-energy at the Fermi level from  
the first crossing diagram (Fig. \ref{Fig:Diagrams}c) is  
\begin{eqnarray*}
\fl \Sigma^{(2c)} (\bk_F, i \omega_n)&=&g_0^2 T \sum_{n^\prime}
         D^{(0)} ( \omega_n -  \omega_{n^\prime}) G^{(0)} (\bk_\mathrm{F},  \omega_{n^\prime}) \times \\
         & & \quad\quad \Gamma^{(1)}_{\bk_\mathrm{F}} ( \omega_{n^\prime}, \omega_n).
\end{eqnarray*}
Performing the Matsubara sum yields 
\begin{eqnarray*}
    \Sigma^{(2c)} (\bk_F, i \omega_n) &=& 
    - \frac{\lambda_\mathrm{m}^2 \Omega^4}{i \omega_n} \bigg[ \frac{1}{ (\Omega^2 + \omega_n^2)} + \mathrm{csch}^2 \left( \frac{\Omega}{2T} \right)\times\\
& & \frac{\Omega^2 (4 \Omega^2 + \omega_n^2) + \omega_n^2 (2 \omega_n^2 - \Omega^2) \cosh \frac{\Omega}{T}}{2 (\Omega^2 + \omega_n^2)^2 (4 \Omega^2 + \omega_n^2)} \bigg].
    \label{CrossingPFS}
\end{eqnarray*}
For temperatures $T \ll \Omega$, the hyperbolic cosecant of $\Omega/2T$ approaches zero and 
the crossing diagram can be approximated as 
\begin{equation}
    \Sigma^{(2c)} (\bk_\mathrm{F}, i \omega_n) = - \frac{\lambda_\mathrm{m}^2 \Omega^4}{ i \omega_n (\Omega^2 + \omega_n^2)}.  
\end{equation}

The above is completely contributed by the elastic term
$\omega_{n'}=\omega_{n}$ in the sum of internal Matsubara frequencies
$\omega_{n'}$.  This shows that the diverging elastic vertex correction
produces a 
self-energy contribution at the Fermi level that diverges as  
$\Sigma(\bk_\mathrm{F},i\omega_n) \sim 1/i\omega_n$. 
This can be understood from the structure of the crossing diagram in Fig. 
\ref{Fig:Diagrams}b. Whenever one of the phonons has a zero frequency,
the total diagram is multiplied by the Greens function $G^{(0)} (\bk_\mathrm{F}, i
\omega_{n}) = 1/i \omega_n$. This contribution is usually suppressed by a
factor $T$; however, here the vertex correction diverges as $1/T$ and we obtain a 
non-vanishing $1/i \omega_n$ contribution.

\end{document}